\documentclass[twocolumn, onecolappendix]{aastex701}

\usepackage{multirow, amsmath}
\usepackage{CJKutf8}
\usepackage[utf8]{inputenc}

\usepackage[multiple]{footmisc}
\usepackage{subcaption}
\received{}
\revised{}
\accepted{}
\submitjournal{ApJ}

\usepackage{soul}
\newcommand\N[1]{{\color{blue} \bf #1}}

\shorttitle{Asymmetries in SN~2023ixf}
\shortauthors{Hsu et al.}

\defcitealias{sdss_2017}{SDSS Collaboration 2017}
\defcitealias{astropy_2018}{Astropy Collaboration 2018}
\defcitealias{Pyraf_2012}{Science Software Branch at STScI 2012}
\begin{document}
\title{Evidence for Asymmetric Ejecta and Circumstellar Material in SN~2023ixf Inferred from Extensive Nebular-phase Observations}

\correspondingauthor{Brian Hsu}
\email{bhsu@arizona.edu}

\newcommand{\LCO}{\affiliation{Las Cumbres Observatory, 6740 Cortona Drive, Suite 102, Goleta, CA 93117-5575, USA}}
\newcommand{\UCSB}{\affiliation{Department of Physics, University of California, Santa Barbara, CA 93106-9530, USA}}
\newcommand{\KITP}{\affiliation{Kavli Institute for Theoretical Physics, University of California, Santa Barbara, CA 93106-4030, USA}}
\newcommand{\UCD}{\affiliation{Department of Physics and Astronomy, University of California, 1 Shields Avenue, Davis, CA 95616-5270, USA}}
\newcommand{\Berkeley}{\affiliation{Department of Astronomy, University of California, Berkeley, CA 94720-3411, USA}}
\newcommand{\WIS}{\affiliation{Department of Particle Physics and Astrophysics, Weizmann Institute of Science, 76100 Rehovot, Israel}}
\newcommand{\OKC}{\affiliation{Oskar Klein Centre, Department of Astronomy, Stockholm University, Albanova University Centre, SE-106 91 Stockholm, Sweden}}
\newcommand{\OAPD}{\affiliation{INAF-Osservatorio Astronomico di Padova, Vicolo dell'Osservatorio 5, I-35122 Padova, Italy}}
\newcommand{\Caltech}{\affiliation{Cahill Center for Astronomy and Astrophysics, California Institute of Technology, Mail Code 249-17, Pasadena, CA 91125, USA}}
\newcommand{\GSFC}{\affiliation{Astrophysics Science Division, NASA Goddard Space Flight Center, Mail Code 661, Greenbelt, MD 20771, USA}}
\newcommand{\UMD}{\affiliation{Joint Space-Science Institute, University of Maryland, College Park, MD 20742, USA}}
\newcommand{\UCB}{\affiliation{Department of Astronomy, University of California, Berkeley, CA 94720-3411, USA}}
\newcommand{\TTU}{\affiliation{Department of Physics, Texas Tech University, Box 41051, Lubbock, TX 79409-1051, USA}}
\newcommand{\STSci}{\affiliation{Space Telescope Science Institute, 3700 San Martin Dr, Baltimore, MD 21218, USA}}
\newcommand{\UT}{\affiliation{University of Texas at Austin, 1 University Station C1400, Austin, TX 78712-0259, USA}}
\newcommand{\IoA}{\affiliation{Institute of Astronomy, University of Cambridge, Madingley Road, Cambridge CB3 0HA, UK}}
\newcommand{\QUB}{\affiliation{Astrophysics Research Centre, School of Mathematics and Physics, Queen's University Belfast, Belfast BT7 1NN, UK}}
\newcommand{\IPAC}{\affiliation{Spitzer Science Center, California Institute of Technology, Pasadena, CA 91125, USA}}
\newcommand{\JPL}{\affiliation{Jet Propulsion Laboratory, California Institute of Technology, 4800 Oak Grove Dr, Pasadena, CA 91109, USA}}
\newcommand{\Southampton}{\affiliation{Department of Physics and Astronomy, University of Southampton, Southampton SO17 1BJ, UK}}
\newcommand{\LANL}{\affiliation{Space and Remote Sensing, MS B244, Los Alamos National Laboratory, Los Alamos, NM 87545, USA}}
\newcommand{\Tsinghua}{\affiliation{Physics Department and Tsinghua Center for Astrophysics, Tsinghua University, Beijing, 100084, People's Republic of China}}
\newcommand{\NAOC}{\affiliation{National Astronomical Observatory of China, Chinese Academy of Sciences, Beijing, 100012, People's Republic of China}}
\newcommand{\Itagaki}{\affiliation{Itagaki Astronomical Observatory, Yamagata 990-2492, Japan}}
\newcommand{\Einstein}{\altaffiliation{Einstein Fellow}}
\newcommand{\Hubble}{\altaffiliation{Hubble Fellow}}
\newcommand{\CfA}{\affiliation{Center for Astrophysics \textbar{} Harvard \& Smithsonian, 60 Garden Street, Cambridge, MA 02138-1516, USA}}
\newcommand{\UA}{\affiliation{Steward Observatory, University of Arizona, 933 North Cherry Avenue, Tucson, AZ 85721-0065, USA}}
\newcommand{\MPA}{\affiliation{Max-Planck-Institut f\"ur Astrophysik, Karl-Schwarzschild-Stra\ss e 1, D-85748 Garching, Germany}}
\newcommand{\DSFP}{\altaffiliation{LSSTC Data Science Fellow}}
\newcommand{\HCO}{\affiliation{Harvard College Observatory, 60 Garden Street, Cambridge, MA 02138-1516, USA}}
\newcommand{\Carnegie}{\affiliation{Observatories of the Carnegie Institute for Science, 813 Santa Barbara Street, Pasadena, CA 91101-1232, USA}}
\newcommand{\TAU}{\affiliation{School of Physics and Astronomy, Tel Aviv University, Tel Aviv 69978, Israel}}
\newcommand{\Edinburgh}{\affiliation{Institute for Astronomy, University of Edinburgh, Royal Observatory, Blackford Hill EH9 3HJ, UK}}
\newcommand{\Birmingham}{\affiliation{Birmingham Institute for Gravitational Wave Astronomy and School of Physics and Astronomy, University of Birmingham, Birmingham B15 2TT, UK}}
\newcommand{\CIERA}{\affiliation{Center for Interdisciplinary Exploration and Research in Astrophysics and Department of Physics and Astronomy, \\Northwestern University, 1800 Sherman Avenue 8th Floor, Evanston, IL 60201, USA}}
\newcommand{\Bath}{\affiliation{Department of Physics, University of Bath, Claverton Down, Bath BA2 7AY, UK}}
\newcommand{\CTIO}{\affiliation{Cerro Tololo Inter-American Observatory, National Optical Astronomy Observatory, Casilla 603, La Serena, Chile}}
\newcommand{\Potsdam}{\affiliation{Institut f\"ur Physik und Astronomie, Universit\"at Potsdam, Haus 28, Karl-Liebknecht-Str. 24/25, D-14476 Potsdam-Golm, Germany}}
\newcommand{\INPE}{\affiliation{Instituto Nacional de Pesquisas Espaciais, Avenida dos Astronautas 1758, 12227-010, S\~ao Jos\'e dos Campos -- SP, Brazil}}
\newcommand{\UNC}{\affiliation{Department of Physics and Astronomy, University of North Carolina, 120 East Cameron Avenue, Chapel Hill, NC 27599, USA}}
\newcommand{\Ohio}{\affiliation{Astrophysical Institute, Department of Physics and Astronomy, 251B Clippinger Lab, Ohio University, Athens, OH 45701-2942, USA}}
\newcommand{\AAS}{\affiliation{American Astronomical Society, 1667 K~Street NW, Suite 800, Washington, DC 20006-1681, USA}}
\newcommand{\MMT}{\affiliation{MMT and Steward Observatories, University of Arizona, 933 North Cherry Avenue, Tucson, AZ 85721-0065, USA}}
\newcommand{\Geneva}{\affiliation{ISDC, Department of Astronomy, University of Geneva, Chemin d'\'Ecogia, 16 CH-1290 Versoix, Switzerland}}
\newcommand{\IUCAA}{\affiliation{Inter-University Center for Astronomy and Astrophysics, Post Bag 4, Ganeshkhind, Pune, Maharashtra 411007, India}}
\newcommand{\CMU}{\affiliation{Department of Physics, Carnegie Mellon University, 5000 Forbes Avenue, Pittsburgh, PA 15213-3815, USA}}
\newcommand{\Columbia}{\affiliation{Simons Junior Fellow, Department of Astronomy, Columbia University, New York, NY 10027-6601, USA}}
\newcommand{\IAIFI}{\affiliation{The NSF AI Institute for Artificial Intelligence and Fundamental Interactions}}
\newcommand{\CFHTC}{\affiliation{Canada-France-Hawaii Telescope Corp., 65-1238 Māmalahoa Highway, Kamuela, HI 96743, USA}}
\newcommand{\IfA}{\affiliation{Institute for Astronomy, University of Hawai'i, 2680 Woodlawn Drive, Honolulu, HI 96822-1839, USA}}
\newcommand{\Princeton}{\affiliation{Department of Astrophysical Sciences, Princeton University, 4 Ivy Lane, Princeton, NJ 08540-7219, USA}}
\newcommand{\Flatiron}{\affiliation{Center for Computational Astrophysics, Flatiron Institute, 162 5th Avenue, New York, NY 10010-5902, USA}}
\newcommand{\GeminiNorth}{\affiliation{Gemini Observatory, 670 North A`ohoku Place, Hilo, HI 96720-2700, USA}}
\newcommand{\STScI}{\affiliation{Space Telescope Science Institute, 3700 San Martin Drive, Baltimore, MD 21218, USA}}
\newcommand{\ICE}{\affiliation{Institute of Space Sciences (ICE, CSIC), Campus UAB, Carrer
de Can Magrans, s/n, E-08193 Barcelona, Spain}}
\newcommand{\IEEC}{\affiliation{Institut d'Estudis Espacials de Catalunya (IEEC), Edifici RDIT, Campus UPC, 08860 Castelldefels (Barcelona), Spain}}
\newcommand{\konkoly}{\affiliation{Konkoly Observatory, HUN-REN Research Center for Astronomy and Earth Sciences, Konkoly Th. M. út 15-17., Budapest, 1121 Hungary; MTA Centre of Excellence}}
\newcommand{\szeged}{\affiliation{Department of Experimental Physics, Institute of Physics, University of Szeged, D\'om t\'er 9, Szeged, 6720 Hungary}}
\newcommand{\UCSD}{\affiliation{Department of Astronomy \& Astrophysics, University of California, San Diego, 9500 Gilman Drive, MC 0424, La Jolla, CA 92093-0424, USA}}
\newcommand{\Monash}{\affiliation{School of Physics and Astronomy, Monash University, Clayton, Victoria 3800, Australia}}
\newcommand{\OzGrav}{\affiliation{OzGrav: The ARC Center of Excellence for Gravitational Wave Discovery, Australia}}
\newcommand{\Adler}{\affiliation{Adler Planetarium, 1300 S. DuSable Lake Shore Dr., Chicago, IL 60605, USA}}
\newcommand{\Keck}{\affiliation{W.~M.~Keck Observatory, 65-1120 M\=amalahoa Highway, Kamuela, HI 96743-8431, USA}}
\newcommand{\Rutgers}{\affiliation{Department of Physics and Astronomy, Rutgers, the State University of New Jersey,\\136 Frelinghuysen Road, Piscataway, NJ 08854-8019, USA}}
\newcommand{\Catalyst}{\altaffiliation{LSST-DA Catalyst Fellow}}
\newcommand{\JHU}{\affiliation{Department of Physics and Astronomy, The Johns Hopkins University, 3400 North Charles Street, Baltimore, MD 21218, USA}}
\newcommand{\SETI}{\affiliation{SETI Institute, 339 Bernardo Ave., Ste. 200, Mountain View, CA 94043, USA}}
 
\author[0000-0002-9454-1742,gname=Brian,sname=Hsu]{Brian~Hsu}
\UA
\email{bhsu@arizona.edu}

\author[0000-0001-5510-2424]{Nathan~Smith}
\UA
\email{nathans@as.arizona.edu}

\author[0000-0002-4924-444X,gname=Azalee,sname=Bostroem]{K.~Azalee~Bostroem}
\affiliation{IPAC, Mail Code 100-22, Caltech, 1200 E.\ California Blvd., Pasadena, CA 91125}
\email{bostroem@ipac.caltech.edu}

\author[0000-0002-0744-0047]{Jeniveve~Pearson}
\UA
\email{jenivevepearson@arizona.edu}

\author[0000-0003-4102-380X]{David~J.~Sand}
\UA
\email{dsand@arizona.edu}

\author[0000-0003-3108-1328]{Lindsey~A.~Kwok}
\CIERA
\email{lindsey.kwok@northwestern.edu}

\author[0000-0003-0123-0062]{Jennifer~E.~Andrews}
\GeminiNorth
\email{jandrews@gemini.edu}

\author[0000-0001-8073-8731]{Bhagya M.~Subrayan}
\UA
\email{bsubrayan@arizona.edu}

\author[0000-0002-0832-2974]{Griffin~Hosseinzadeh}
\UCSD
\email{ghosseinzadeh@ucsd.edu}

\author[0000-0002-4022-1874]{Manisha~Shrestha}
\Monash \OzGrav 
\email{manisha.shrestha@monash.edu}

\author[orcid=0000-0003-4175-4960]{Conor Ransome}
\UA
\email{cransome@arizona.edu}

\author[0000-0002-1895-6639]{Moira Andrews}
\LCO\UCSB
\email{mandrews@lco.global}

\author[0000-0003-0528-202X]{Collin~T.~Christy}
\UA
\email{collinchristy@arizona.edu}

\author[0000-0002-7937-6371]{Yize Dong \begin{CJK*}{UTF8}{gbsn}(董一泽)\end{CJK*}}
\CfA
\email{yize.dong@cfa.harvard.edu}

\author[0000-0003-4914-5625]{Joseph Farah}
\LCO\UCSB
\email{jfarah@lco.global}

\author[0000-0003-3460-0103]{Alexei V. Filippenko}
\Berkeley
\email{afilippenko@berkeley.edu}

\author[0000-0003-4537-3575]{Noah~Franz}
\UA
\email{nfranz@arizona.edu}

\author[0000-0003-1012-3031]{Jared~A.~Goldberg}
\affil{Department of Physics and Astronomy, Michigan State University, East Lansing, MI 48824, USA}
\email{goldstar@msu.edu}

\author[0000-0003-0209-9246]{Estefania~Padilla~Gonzalez}
\JHU
\email{epadill7@jh.edu}

\author[0000-0003-2375-2064]{Claudia~P.~Guti\'errez }
\ICE\IEEC
\email{cgutierrez@ice.csic.es}

\author[0000-0003-2744-4755]{Emily~Hoang}
\UCD
\email{emthoang@ucdavis.edu}

\author[0000-0003-4253-656X]{D.~Andrew~Howell}
\LCO\UCSB
\email{ahowell@lco.global}

\author[0000-0001-8738-6011]{Saurabh~W.~Jha}
\Rutgers
\email{saurabh@physics.rutgers.edu}

\author[0000-0002-8770-6764]{R\'eka~K\"onyves-T\'oth}
\konkoly
\email{konyvestoth.reka@csfk.org}

\author[0000-0001-9589-3793]{Michael~Lundquist}
\Keck
\email{mlundquist@keck.hawaii.edu}

\author[0000-0001-5807-7893]{Curtis~McCully}
\LCO\UCSB
\email{cmccully@lco.global}

\author[0009-0008-9693-4348]{Darshana~Mehta}
\UCD
\email{ddmehta@ucdavis.edu}

\author[0000-0002-7015-3446]{Nicolas~E.~Meza Retamal}
\UCD
\email{nemezare@ucdavis.edu}

\author[0000-0001-9570-0584]{Megan~Newsome}
\UT
\email{newsome.megane@gmail.com}

\author[0000-0002-7352-7845]{Aravind~P.~Ravi}
\UCD
\email{apazhayathravi@ucdavis.edu}

\author[0000-0003-3643-839X]{Jeonghee~Rho}
\SETI
\email{jrho@seti.org}

\author[0000-0003-0794-5982]{Giacomo~Terreran}
\Adler
\email{gterreran@adlerplanetarium.org}

\author[0000-0001-8818-0795]{Stefano~Valenti}
\UCD
\email{valenti@ucdavis.edu}

\author[0000-0002-4951-8762]{Sergiy Vasylyev}
\UCSD
\email{svasylyev@ucsd.edu}

\author[0000-0002-7334-2357]{X.-F.~Wang}
\Tsinghua
\email{wang_xf@mail.tsinghua.edu.cn}

\author[orcid=0009-0006-7296-728X]{Kathryn Wynn}
\LCO \UCSB 
\email{kwynn@ucsb.edu}

\author[0000-0002-6535-8500]{Yi Yang}
\Tsinghua
\email{yi_yang@mail.tsinghua.edu.cn}

\begin{abstract}
We present extensive optical and near-infrared (NIR) observations of the nearby Type II supernova (SN II) 2023ixf in the nebular phase from $+89$ days to $+749$ days after explosion, supplemented with NIR and mid-infrared (MIR) spectroscopy from the James Webb Space Telescope. The H$\alpha$ emission profile shows complex evolution, with the emergence of high-velocity components consistent with the outer ejecta interacting with extended, low-density circumstellar material (CSM). We find that the H$\alpha$ profile at an intermediate epoch (around $+375$ d) can be reconstructed by scaling an earlier decay-powered component and a later-phase shock-powered component, which revealed an additional intermediate-width component. This is consistent with the ejecta crashing into the initially aspherical dense CSM that has been swept-up by the forward shock. In the NIR, we find double-peaked emission from \ion{Mg}{1} $1.504\ {\rm \mu m}$, \ion{Na}{1} $2.206\ {\rm \mu m}$, and [\ion{Ni}{1}] $3.12\ {\rm \mu m}$ between $+200$ d and $+374$ d, consistent with an asymmetric distribution of Ni-rich material that heats the ejecta inhomogeneously. We posit a disk-like CSM geometry and an ejecta geometry in which at least two large Ni-rich plumes lead to the observed line-profile diversity.
\end{abstract}

\section{Introduction}
\label{sec:intro}

Core-collapse supernovae (CCSNe) mark the explosive demise of massive stars ($\gtrsim 8\ M_{\odot}$; \citealt{Woosley_Weaver_1986}) and represent one of the most fundamental processes that shape the Universe on every scale.
Understanding the explosion mechanism and nucleosynthetic yields of CCSNe are longstanding goals in supernova (SN) science. 
Modern three-dimensional (3D) simulations reveal that many aspects of CCSNe are generically asymmetric\footnote{Throughout this work, we distinguish between {\it aspherical} and {\it asymmetric} geometries. We use ``aspherical" to denote deviations from spherical symmetry that may still preserve axisymmetry (e.g., oblate or disk-like configurations), while ``asymmetric" refers more generally to configurations that lack any axisymmetry.}, including the pre-explosion stellar envelopes (\citealt{Chiavassa_2009,Chiavassa_2010,Goldberg_2022a,Goldberg_2022b,Goldberg_2026}), stellar cores (\citealt{Fields_2020,Fields_2021,Yadav_2020,Rizzuti_2024}), circumstellar material (CSM; \citealt{Tsang_2022,Orlando_2024,Ma_2025}), and explosion mechanisms (\citealt{Janka_2012,Wongwathanarat_2015,Wongwathanarat_2017,Burrows_2019,Burrows_2020,Muller_2017,Vartanyan_2019,Vartanyan_2022,Vartanyan_2025b,Vartanyan_2025,Sandoval_2021}). 
These large-scale asymmetries have been confirmed observationally via a multitude of approaches, including strong spectropolarimetric signatures (e.g., \citealt{Leonard_2001,Leonard_2006}), resolved imaging of SN~1987A (\citealt{Larsson_2019}) and SN remnant Cassiopeia A (\citealt{Milisavljevic_2024}), and dynamical kicks measured in neutron stars (\citealt{Hobbs_2005}).

Despite major advances in self-consistent 3D simulations, the detailed distribution of material in CCSN ejecta remains a central unresolved problem. 
Most state-of-the-art 3D simulations of the core-collapse mechanism do not calculate radiation hydrodynamics beyond $\sim1$ s after core bounce, after which some 3D processes continue to synthesize heavy elements (\citealt{Wang_Burrows_2024}). 
While some effort has been made to extend simulations out to late times and incorporate all 3D processes that affect the final nucleosynthetic yields  (\citealt{van_Baal_2023,Wang_Burrows_2024,van_Baal_2025,Vartanyan_2025b}), the number of 3D simulations is still far from sufficient to make conclusive remarks on the ejecta structure and elemental distribution of CCSNe. 
In this context, nebular observations of CCSNe are particularly useful, as they reveal crucial information about the composition, geometry, kinematics, and ionization of the ejected material. 

At epochs $\gtrsim100$ days post-explosion, the opacity in the SN ejecta drops significantly, and the spectra are dominated by a combination of permitted and forbidden emission lines powered by high-energy photons from the radioactive decay chain of $^{56}$Ni$\rightarrow$ $^{56}$Co $\rightarrow$ $^{56}$Fe (\citealt{Nadyozhin_1994}). These nebular lines help probe the elemental distribution and asymmetries in the ejecta through line strengths, shapes, and widths. Historically, optical and near-infrared (NIR) emission lines like H$\alpha$, [\ion{O}{1}] $\lambda\lambda$6300, 6364, [\ion{Ca}{2}] $\lambda\lambda7291$, 7324, [\ion{Fe}{2}] 1.257, 1.644 $\mu$m, and \ion{Mg}{1} 1.504, 1.711 $\mu$m have been used to inform our understanding of the structure and chemical distribution in CCSN ejecta (e.g., \citealt{Elmhamdi_2003,Modjaz_2008,Taubenberger_2009,Maguire_2010,Milisavljevic_2010,Jerkstrand_2012,Jerkstrand_2014,Jerkstrand_2015b,Bose_2019,Szalai_2019}). 
Persistent asymmetries observed in H$\alpha$ and [\ion{O}{1}] $\lambda\lambda$6300, 6364 have been highlighted as direct evidence for ejecta asymmetries and the presence of dust (\citealt{Andrews_2010,Kuncarayakti_2015,Bevan_2017,Bevan_2018,Fang_2022,Fang_2024}). 
Spherically symmetric one-dimensional (1D) models have also been used to infer He core mass, which is a direct tracer of the progenitor's initial mass (\citealt{Jerkstrand_2012}). 

Optical and NIR emission lines are subject to significant blending with nearby species, telluric absorption (if taken from ground-based facilities), and attenuation from any newly formed dust within the ejecta, complicating the interpretation of line widths and profile shapes. 
Synergy of optical and NIR spectra with mid-infrared (MIR) observations, where emission lines are more isolated and less attenuated, mitigates  these limitations. 
Such observations of CCSNe, however, are few and far between, limited to very nearby events like SN~1987A observed with the Kuiper Airborne Observatory (\citealt{Wooden_1993}), as well as SN~2004dj (\citealt{Szalai_2011}), SN~2004et (\citealt{Kotak_2009}), and SN~2005af (\citealt{Kotak_2006}) observed with the Spitzer Space Telescope. Now, the superior sensitivity and resolution of the James Webb Space Telescope (JWST), in conjunction with high-quality, high-cadence ground-based optical and NIR observations, provide the necessary information to infer the structure of CCSN ejecta in unprecedented detail (\citealt{Larsson_2023,Milisavljevic_2024,Dessart_2025c,Medler_2025,Medler_2026}).

In this work, we present new optical and NIR observations of the nearby Type II SN~2023ixf during its nebular phase from $+89$ days to $+749$ days after explosion and analyze its near-concurrent optical-to-MIR spectra to constrain the ejecta geometry and elemental distribution. 
SN~2023ixf is one of the closest hydrogen-rich SNe (SNe II) observed in the past few decades (\citealt{Perley_2023}), discovered on 2023 May 19 17:27:15.00 UTC (\citealt{Itagaki_2023}) in the Pinwheel galaxy (M101). 
To date, almost $\sim 100$ papers have been published on various aspects of SN~2023ixf, including the dust-enshrouded red supergiant (RSG) progenitor and its local environment (\citealt{Dong_2023,Liu_2023,Jencson_2023,Kilpatrick_2023,Niu_2023,Neustadt_2024,Pledger_Shara_2023,Soker_2023,Soraisam_2023,Fuller_2024,Ransome_2024,Qin_2024,Van_Dyk_2024,Xiang_2024,Rest_2025}), multiwavelength observations after shock breakout (\citealt{Berger_2023,Bostroem_2023, Bostroem_2024,Flinner_2023,Grefenstette_2023,Hiramatsu_2023,Hosseinzadeh_2023, Jacobson-Galan_2023,Sgro_2023,Smith_2023,Teja_2023,Yamanaka_2023,Zhang_2023,Chandra_2024,Gong_2024,Li_2024,St-Onge_2024,Van_Dyk_2024b,Yang_2024,Zimmerman_2024,Dickinson_2025,Iwata_2025,Jacobson-Galan_2025,Nayana_2025,DerKacy_2026,Ragosta_2026}), light-curve modeling (\citealt{Bersten_2024,Martinez_2024,Hsu_2025,Hu_2025,Moriya_Singh_2024,Fang_2025b,Forde_2025,Kozyreva_2025, Vinko_2025,Laplace_2026,Utrobin_2026}), spectropolarimetric observations (\citealt{Vasylyev_2023,Vasylyev_2026,Singh_2024,Shrestha_2025}), multimessenger searches (\citealt{Guetta_2023, Kheirandish_2023,Marti-Devesa_2024,Ravensburg_2024, Sarmah_2024,Abac_2025,Cosentino_2025,Kimura_2025}), nebular emission-line profiles (\citealt{Ferrari_2024,Singh_2024,Singh_2026,Bostroem_2025,Folatelli_2025,Kumar_2025,Li_2025,Michel_2025,Zheng_2025}), molecular and thermal dust emission (\citealt{Medler_2025, Park_2025,Singh_2026}), and alternative mechanisms (\citealt{Reynoso_2024,Soker_2025}). 
For the sake of brevity, we do not provide a detailed summary here and refer readers to \cite{Jacobson-Galan_2025b} for a comprehensive overview.

We present our observations and data reduction in Section~\ref{sec:obs}. 
Section~\ref{sec:spectral_evo} describes the spectral evolution of several prominent and interesting nebular emission lines.
We analyze the H$\alpha$ profile evolution in Section~\ref{sec:Halpha} and empirically model selected emission lines in Section~\ref{sec:line_profile_analysis}.
In Sections~\ref{sec:discussion} and Section~\ref{sec:conclusions}, respectively, we discuss our findings and present our conclusions.

\section{Observations}
\label{sec:obs}

Here we adopt the Cepheid distance to the Pinwheel Galaxy of $d_L=6.9\pm0.15$ Mpc ($\mu=29.178\pm 0.041$ mag; \citealt{Leavitt_1908,Riess_2022}).
All observations are corrected for foreground Milky Way extinction of $E(B-V)_{\rm MW}=0.0077$ mag (\citealt{Schlafly_Finkbeiner_2011}) and host-galaxy extinction of $E(B-V)_{\rm host}=0.031\pm0.006$ mag (\citealt{Lundquist_2023,Smith_2023}), using the {\tt astropy} (\citetalias{astropy_2018}) implementation of the \cite{Gordon_2023} extinction law.
We adopt MJD 60082.788 as the explosion date, following the analysis of \cite{Li_2024}. 
New photometry and spectroscopy presented in this paper will be available as data behind figures and uploaded to the Open mulTiwavelength Transient Event Repository\footnote{\url{https://otter.idies.jhu.edu}} (OTTER; \citealt{Franz_2026}).

\subsection{Optical Photometry} 
\label{sec:photometry}

\begin{figure*}[t!]
    \centering
    \includegraphics[width=\textwidth]{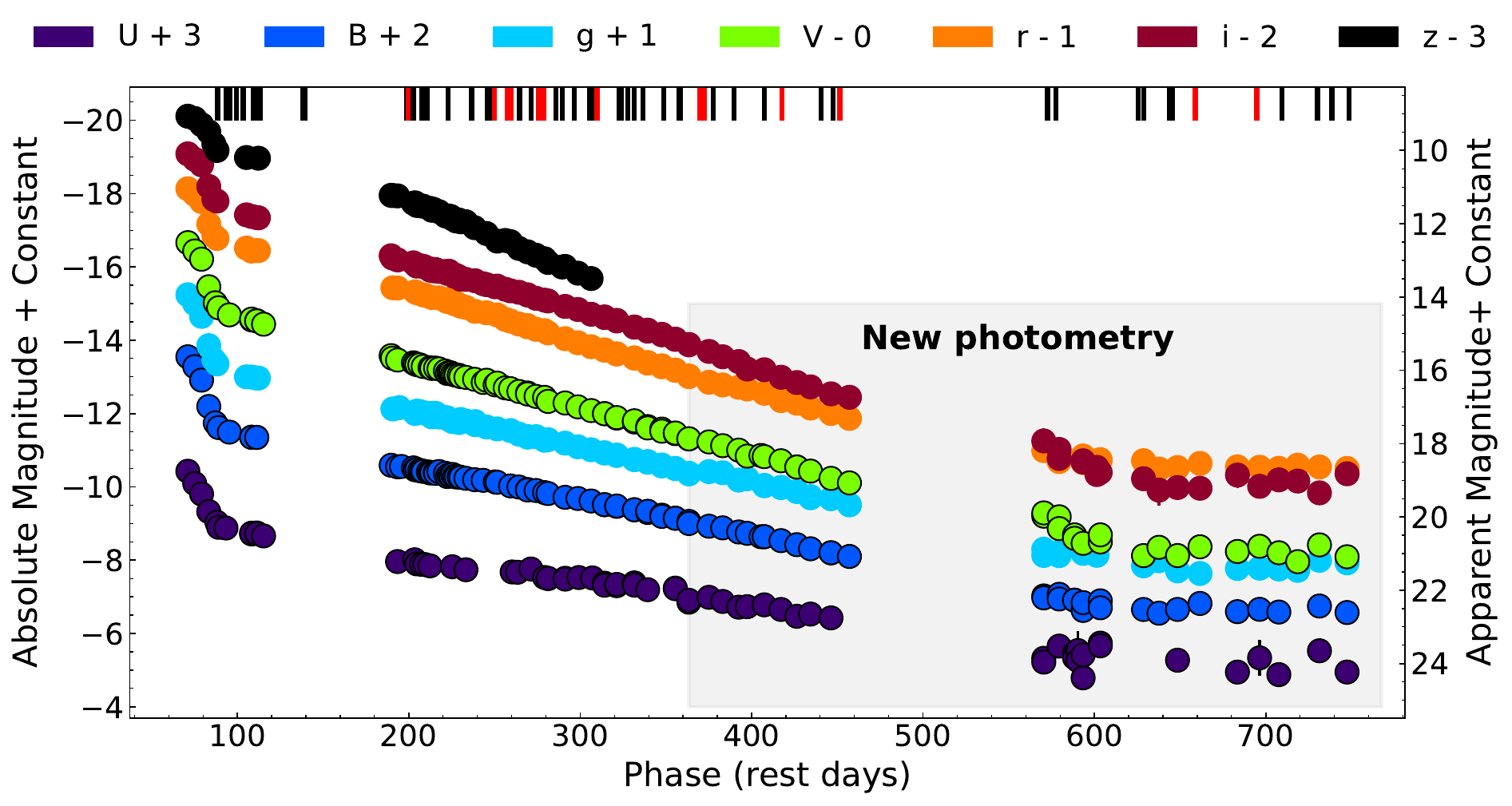}
    \caption{Optical ($UBgVriz_s$) light curves of SN~2023ixf from LCO, corrected for Milky Way and host extinctions. The error bars of most observed magnitudes are smaller than the marker size ($\lesssim 0.1$ mag). 
    The LCO photometry up to the first year ($\sim 360$ days) of SN~2023ixf's evolution was previously presented by \cite{Hsu_2025}. 
    New photometry presented in this work is highlighted in the gray shaded box.
    The times of the spectral epochs are marked at the top with vertical black lines (optical) and red (NIR).\\(The photometry used to create this figure is available as data behind the figure.)}
    \label{fig:lc}
\end{figure*}

In Figure~\ref{fig:lc}, we show follow-up $UBgVriz_s$ photometry from Las Cumbres Observatory’s robotic 1~m telescopes \citep{Brown_2013} obtained as part of the Global Supernova Project (GSP) collaboration (\citealt{Howell_2017}).
Photometric reductions were carried out using {\tt lcogtsnpipe} (\citealt{Valenti_2016}), a {\tt PyRAF}-based image reduction pipeline that utilizes a standard point-spread function fitting procedure to measure instrumental magnitudes.
{\it UBV} magnitudes were calibrated to stars in the L92 standard fields of \cite{Landolt_1983,Landolt_1992} observed on the same night with the same telescopes and {\it griz$_{\rm s}$} magnitudes were calibrated to the Sloan Digital Sky Survey catalog (\citetalias{sdss_2017}).
More details on the observations, data reduction, and photometry for SN~2023ixf up to 2024 May 16 (MJD = 60446; $\sim$ 1 yr after discovery) can be found in \cite{Hsu_2025}.

\subsection{Optical Spectroscopy} 
\label{sec:opt_spectroscopy}


In this section, we briefly describe the telescopes and instruments used to obtain our optical spectra, as well as their respective reduction processes.

\begin{itemize}
\item {\bf FTN 2~m:} We obtained 18 epochs of optical spectroscopy with the low-resolution cross-dispersed FLOYDS spectrograph (\citealt{Brown_2013}) on Las Cumbres Observatory's 2~m Faulkes Telescope North (FTN) at Haleakalā Observatory (Hawai'i, USA) as part of the Global Supernova Project collaboration (\citealt{Howell_2017}). All observations were taken with a $2.0\arcsec$-wide slit and a  235 lines mm$^{-1}$ grating covering a spectral range of $\sim3200-10,000\ {\rm \AA}$ with a resolution of $R\approx 400$--700 depending on the wavelength. 1D spectra were extracted, reduced, and calibrated following standard procedures using the FLOYDS pipeline (\citealt{Valenti_2014}). 
\item {\bf MMT 6.5~m:} Multiple epochs of spectroscopy were obtained with the 6.5~m MMT telescope, with either the Blue Channel (6 epochs) spectrograph (\citealt{Angel_1979,Schmidt_1989}) or Binospec (\citealt{Fabricant_2019}; 8 epochs). The MMT/Blue Channel observations were taken with a $1.0\arcsec$ slit, two of which using the 1200 lines mm$^{-1}$ grating ($R\approx3340$) covering a range of $\sim5700-7000\ {\rm \AA}$, and one using the 300 lines mm$^{-1}$ grating ($R\approx740$) covering a range of $\sim3600-9000\ {\rm \AA}$. Standard reductions for MMT/Blue Channel observations were carried out using {\tt IRAF} (\citealt{Tody_1986}), including bias subtraction, flat-fielding, and optical extraction of the spectra. Flux calibrations were achieved using spectrophotometric standards observed at an airmass similar to that of each science frame, and the resulting spectra were median combined into a single 1D spectrum for each epoch. The MMT/Binospec observations were taken with a $1.0\arcsec$ slit using the 270 lines mm$^{-1}$ grating ($R\approx1340$) covering a range of $\sim3850-9150\ {\rm \AA}$. All data were reduced either using the Binospec pipeline (\citealt{Kansky_2019}), which includes an internal flux calibration into relative flux units from throughput measurements of spectrophotometric standard stars.
\item {\bf LBT 2$\times$8.4~m:} Eight epochs of spectroscopy were observed with the Multi-Object Double Spectrographs (MODS; \citealt{Byard_2000,Pogge_2010}) on the 2$\times$8.4~m Large Binocular Telescope (LBT). The data were obtained in dichroic mode using both the G400L ($3500–5900\ {\rm \AA}$) and the G670L ($5400–10,000\ {\rm \AA}$) gratings on the blue and red channels on each of the two identical MODS1 and MODS2 spectrographs, achieving a resolution of $R\approx2000$. Raw data were first bias and flat-field corrected using the {\tt modsCCDred} package (\citealt{Pogge_2019a,Pogge_2019b}), then extracted and flux calibrated using {\tt IRAF} (\citealt{Tody_1986}).
\item {\bf Keck I 10~m:} Three spectra were taken with the Low-Resolution Imaging Spectrometer (LRIS; \citealt{Oke_1995}) on the 10 m Keck I telescope and reduced using the {\tt LPipe} package (\citealt{Perley_2019}). 
\item {\bf Bok 2.3~m:} Five epochs of spectroscopy were taken with the Boller \& Chivens (B$\&$C) spectrograph on the Bok 2.3~m telescope with a $1.5\arcsec$ slit using the 300 lines mm$^{-1}$ grating ($R\approx700$) covering a range of $\sim4000-9000\ {\rm \AA}$. Standard reductions were carried out using {\tt IRAF} (\citealt{Tody_1986}) in a similar manner as MMT/Blue Channel observations.
\item {\bf Lick 3~m:} Three epochs of optical spectra were obtained with the Kast spectrograph on the 3~m Shane reflector at Lick Observatory. Kast spectra were reduced using standard {\tt IRAF} (\citealt{Tody_1986})/{\tt PyRAF} (\citetalias{Pyraf_2012}) and {\tt Python} routines for bias/overscan subtractions and flat-fielding. 
\item {\bf Gemini-N 8.1~m:} Lastly, one spectrum was obtained with the Gemini Multi-Object Spectrograph (GMOS; \citealt{Hook_2004, Gimeno_2016}) on the 8.1-m Gemini North Telescope using the B480 grating. Data were reduced using the Data Reduction for Astronomy from Gemini Observatory North and South ({\tt DRAGONS}; \citealt{Labrie_2023}) reduction package with the recipe for GMOS long-slit reductions. This includes bias correction, flat-fielding, wavelength calibration, and flux calibration. 
This spectrum was previously published by \cite{Bostroem_2025}.
\end{itemize}

To account for varying seeing conditions (\citealt{Filippenko_1982}), we scale the fluxes of our spectra to place them on an absolute flux scale using the LCO photometry with the Light Curve Fitting package (\citealt{Hosseinzadeh_2023_lc}). 
At any given spectral epoch, the photometric calibration process converts the interpolated photometry to fluxes at the effective wavelength of each filter and convolves the corresponding spectrum with the transmission function of each filter. 
The ratios of fluxes from the spectrum and photometry are then fit with a first-order polynomial to derive wavelength-dependent correction factors that are applied to the original spectrum. 
A complete log of optical spectroscopy can be found in Table~\ref{tab:opt_spec_log}.

\subsection{NIR Spectroscopy} 
\label{sec:nir_spectroscopy}

Similar to Section~\ref{sec:opt_spectroscopy}, we briefly describe the telescopes and instruments used to obtain our NIR spectra, as well as their respective reduction processes.

\begin{itemize}
\item {\bf IRTF 3.2~m:} Four NIR spectra of SN~2023ixf were observed with the NASA InfraRed Telescope Facility (IRTF) with the SpeX spectrograph (\citealt{Rayner_2003}). Two observations were taken with the short cross-dispersion (SXD) mode and two with the Prism mode, using a $0.8\arcsec$ slit with an ABBA dithering pattern. The associated flat-field and comparison-lamp observations were taken right after the science observation cycles. The IRTF/SpeX data were reduced with Spextool (\citealt{Cushing_2004}) and the output was telluric-corrected using a standard A0\,V star observed at a similar airmass adjacent to the science target, following the prescription of \cite{Vacca_2003}.
The $+279$ d spectrum was previously published by \cite{Park_2025}.
\item {\bf MMT 6.5~m:} We obtained three sets of NIR spectra using the MMT and Magellan Infrared Spectrograph (MMIRS; \citealt{McLeod_2012}) with a $1.0\arcsec$ slit. 
The MMIRS data were manually reduced using the MMIRS pipeline (\citealt{Chilingarian_2015}), then the 1D spectral outputs were telluric- and absolute-flux-corrected following the method described by \citet{Vacca_2003} with the \verb|XTELLCOR_GENERAL| tool (part of Spextool package; \citealt{Cushing_2004}) using a standard A0\,V star observed at a similar airmass and time. The $+200$ d and $+258$ d spectra were previously published by \cite{Park_2025}.
\end{itemize}

A complete log of NIR spectroscopy is listed in Table \ref{tab:nir_spec_log}. 

\begin{figure*}[t!]
    \centering
    \includegraphics[width=\textwidth]{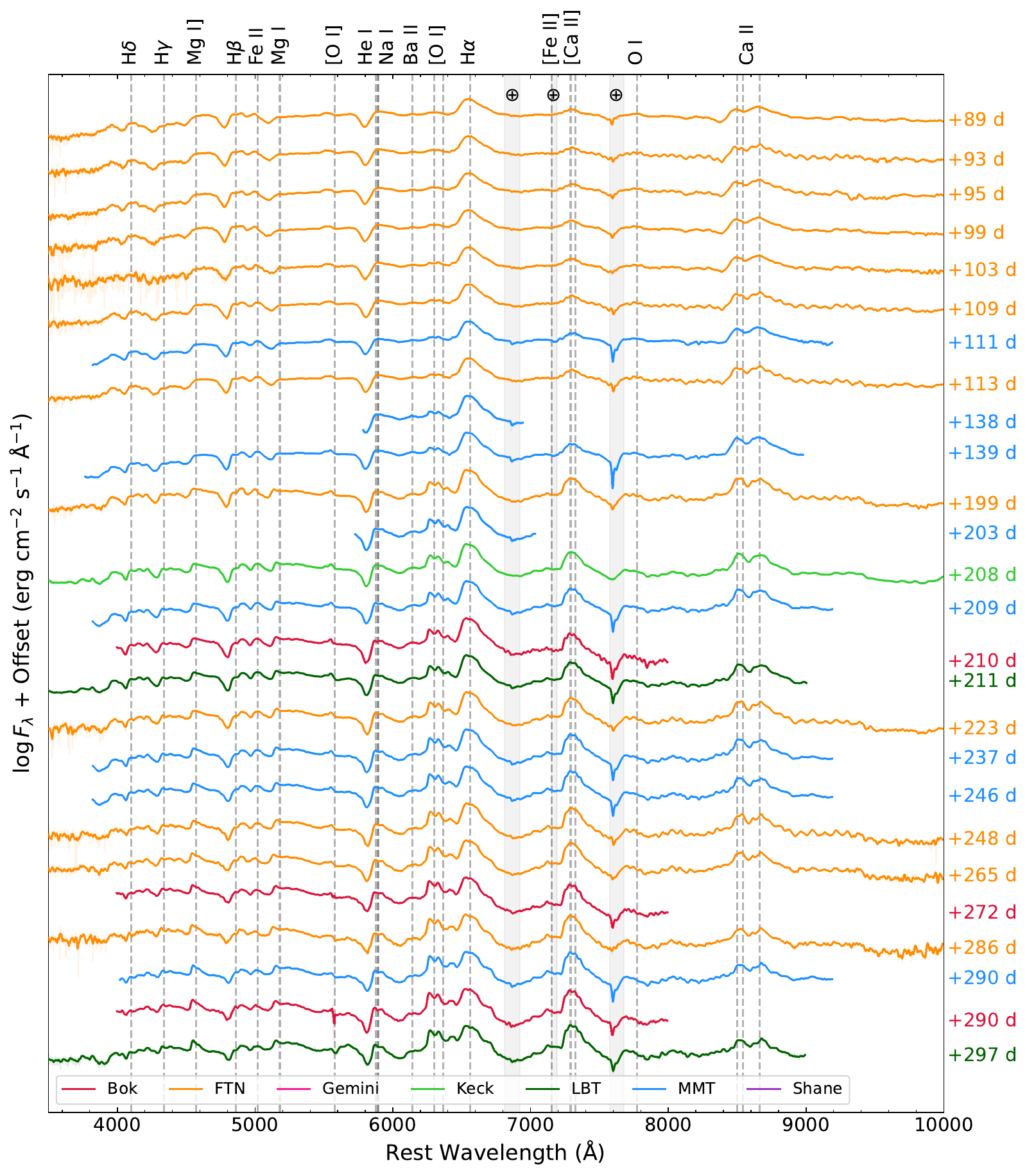}
    \caption{Optical spectra of SN~2023ixf from $+89$ days to $+749$ days after explosion, corrected for $E(B-V)_{\rm tot}=0.0387$ mag and calibrated to optical photometry in Figure~\ref{fig:lc}. 
    Each telescope is denoted by a different color.
    Some spectra have been smoothed with a Savitzky–Golay filter (\citealt{Savitzky_Golay_1964}) to reduce noise for clarity, with the unsmoothed spectra displayed at lower opacity. 
    Shaded gray regions indicate atmospheric absorption bands. 
    Prominent spectral features of SNe II commonly seen in the nebular phase are labeled with vertical dashed lines at their respective rest-frame wavelengths. \\
    (The spectra used to create this figure are available as data behind the figure.)}
    \label{fig:optical_sequence}
\end{figure*}

\begin{figure*}[t!]
    \centering
    \figurenum{2}
    \includegraphics[width=\textwidth]{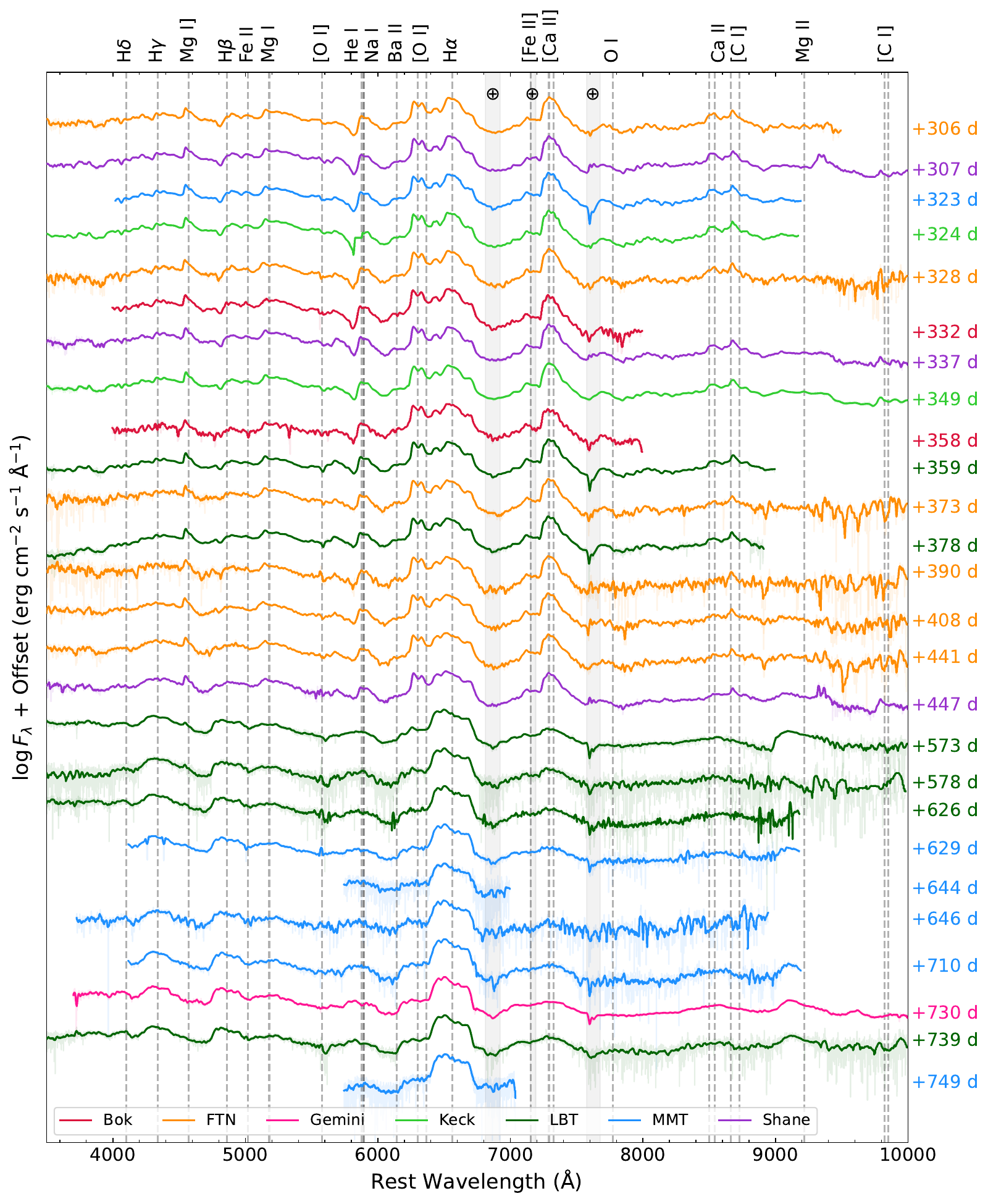}
    \caption{({\it continued})}
\end{figure*}

\subsection{Supplemental Data} 
\label{sec:sup_data}

To extend the wavelength coverage of our spectroscopic dataset into the MIR regime and fill in missing NIR epochs, we additionally supplement our dataset with NIR and MIR spectra taken with JWST from \cite{Medler_2025} and various ground-based NIR spectra from \cite{Park_2025} and \cite{Jacobson-Galan_2025} from WISeREP\footnote{\url{https://www.wiserep.org}} (\citealt{wiserep_2012}). 
We refer readers to \cite{Medler_2025} for details on the JWST spectra taken with the Near-Infrared Spectrograph (NIRSpec; \citealt{Boker_2022}) and Mid-Infrared Instrument in Low Resolution Spectroscopy mode (MIRI/LRS; \citealt{Kendrew_2015}).
As the MIRI/LRS wavelength calibration comes with significant uncertainties, we follow \cite{Kwok_2025a} and correct the wavelength using the equation in their Appendix A. 
Quantitatively, this correction affects shorter wavelengths ($\sim 5-9\ {\rm \mu m}$) the most, which greatly dictates the measured parameters for crucial lines like [\ion{Ni}{2}] $6.636\ {\rm \mu m}$ and [\ion{Ar}{2}] $6.985\ {\rm \mu m}$.

\section{Spectral Evolution}
\label{sec:spectral_evo}

The nebular-phase spectral evolution of SN~2023ixf is shown up to $+749$ d in the optical in Figure~\ref{fig:optical_sequence} and up to $+695$ d in the NIR in Figure~\ref{fig:NIR_sequence}.
All spectra have been shifted to rest-frame wavelengths assuming a redshift of $z=0.000804$ for M101 (\citealt{deVaucouleurs_1995}), and all spectra have been corrected for both Milky Way and host-galaxy extinction. 
We identify emission lines guided by the radiative-transfer models of \cite{Dessart_2025b}, \cite{Dessart_2025c}, and the National Institute of Standards and Technology Atomic Spectra Database\footnote{\url{https://physics.nist.gov/asd}} (\citealt{Kramida2024}). 
The nebular spectra exhibit many of the emission features commonly observed in nebular-phase SNe II, including transitions from H, He, intermediate-mass elements (IMEs; e.g., O, Mg, Ca), and iron-group elements (IGEs; e.g., Fe, Co, Ni), many of which have already been reported in previous studies (e.g., \citealt{Ferrari_2024,Singh_2024,Park_2025,Medler_2025}).
In the following section, we highlight the most prominent and newly identified lines in our data. 
Note that we only focus on emission from atomic and ionic transitions here and ignore molecular or dust emission. 
We show the line-profile evolution of selected lines in Figure~\ref{fig:line_profiles_time}.

\subsection{Hydrogen and Helium: Emergence of Shock-Powered Emission}
\label{sec:hydrogen_helium}

Emission from H dominates the nebular spectra of SN~2023ixf, with a particularly complicated evolution in the H$\alpha$ line profile (see Figure~\ref{fig:line_profiles_time}). 
Between $+89$ d and $+139$ d, H$\alpha$ displays a broad, asymmetric, and blueshifted emission profile. 
In SNe II, blueshifted H$\alpha$ is commonly interpreted as a consequence of optical-depth effects associated with steep density gradients in the expanding material (\citealt{Dessart_Hillier_2005,Anderson_2014}) or dust attenuation (\citealt{Lucy_1989,Bevan_Barlow_2016,Bevan_2017}). 
In the former case, the H$\alpha$ emission peak is expected to migrate toward zero velocity as the ejecta continue to expand. 
In SN~2023ixf, however, we observe a persistently blueshifted H$\alpha$ line profile, in addition to multiple peaks and a slightly slanted shape, which has been suggested as evidence for asymmetric ejecta of other SNe in the absence of dust (e.g., \citealt{Utrobin_1995,Utrobin_2021,Elmhamdi_2003,Chugai_2005,Tomasella_2013,Bose_2015,Andrews_2019,Bose_2019}).

\setcounter{figure}{2}
\begin{figure*}[t!]
    \centering
    \includegraphics[width=\textwidth]{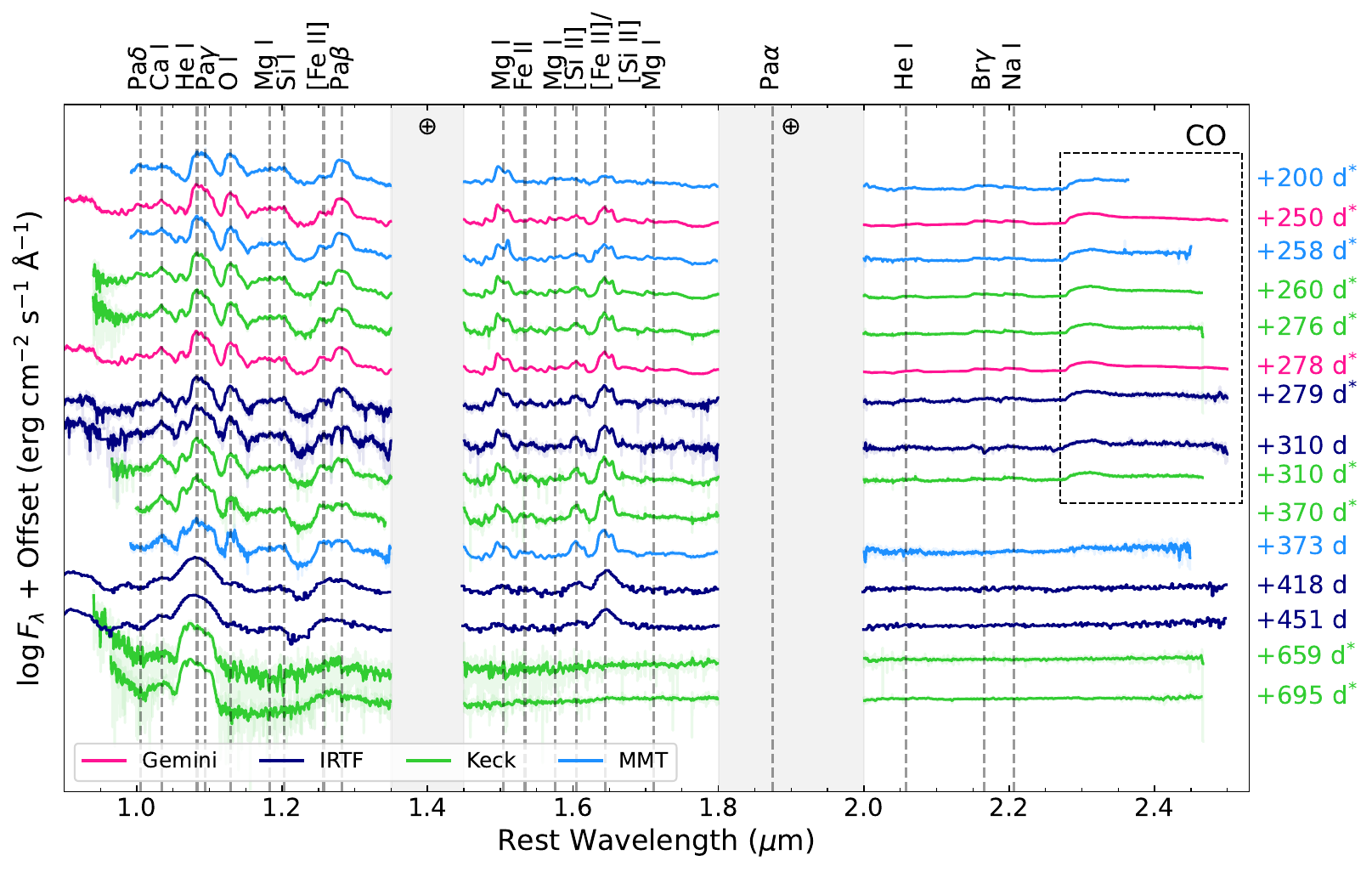}
    \caption{NIR spectra of SN~2023ixf from $+200$ days to $+695$ days after explosion, corrected for $E(B-V)_{\rm tot}=0.0387$ mag. 
    Spectra with an asterisk are either published in \cite{Park_2025} or supplemented from previous studies of SN~2023ixf (\citealt{Park_2025,Jacobson-Galan_2025}) to complete the temporal coverage relative to optical observations. 
    Each telescope is denoted by a different color.
    Some spectra have been smoothed with a Savitzky–Golay filter (\citealt{Savitzky_Golay_1964}) to reduce noise for clarity, with the unsmoothed spectra displayed at lower opacity. 
    Shaded gray regions indicate strong telluric absorption regions. 
    Prominent spectral features of SNe II commonly seen in the nebular phase are labeled with vertical dashed lines at their respective rest-frame wavelengths. 
    The wavelength region of the first overtone CO emission is marked in the dashed black box. 
    We note that Pa$\alpha$ falls in the telluric region.\\
    (Unpublished spectra used to create this figure are available as data behind the figure.)}
    \label{fig:NIR_sequence}
\end{figure*}

At $+199$ d, a blue flux excess appears at approximately $-7500\ {\rm km\ s^{-1}}$, earlier than a similar feature first reported in the $+259$ d spectrum by \cite{Ferrari_2024}, who marked it as unidentified. 
Multiple works  have since interpreted this evolving feature as shock-powered emission from the fast outer edge of H-rich ejecta interacting with low-density CSM sitting farther out from SN~2023ixf's progenitor ($\sim 2\times10^{16}$ cm by $+199$ d, assuming the outermost ejecta are expanding at a velocity of $13,500\ {\rm km\ s^{-1}}$ based on the high-velocity absorption feature in H$\alpha$; \citealt{Singh_2024, Folatelli_2025,Jacobson-Galan_2025,Kumar_2025, Li_2025, Michel_2025,Zheng_2025}). 
This interpretation is consistent with late-time X-ray, ultraviolet (UV), and radio observations of SN~2023ixf (\citealt{Timmerman_2024,Bostroem_2025,Iwata_2025,Jacobson-Galan_2025, Nayana_2025}) and radiative-transfer models (\citealt{Dessart_2023, Dessart_2026}). 

\begin{figure*}[t!]
    \centering
    \includegraphics[width=.9\textwidth]{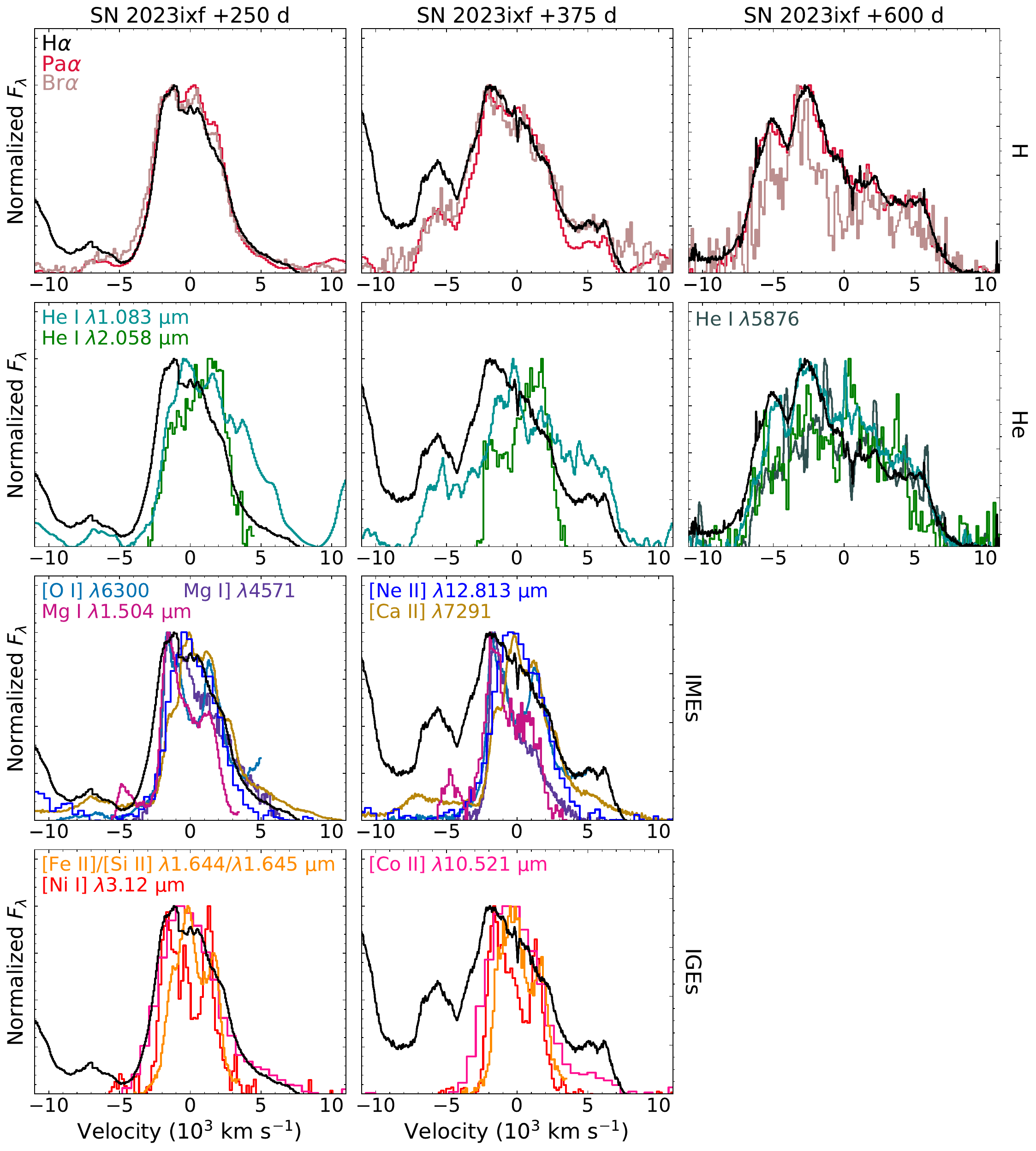}
    \caption{Line-profile comparison of several prominent emission features in SN~2023ixf at three different epochs, separated into four categories. 
    A linear pseudocontinuum is subtracted from each line profile before normalizing to the emission peak. 
    As we do not have optical or NIR spectra around $+600$ d, we use the $+626$ d LBT/MODS spectrum and the $+659$ d Keck/NIRES spectrum.
    This should not affect the line-profile comparison too much, as most emission arises from CSM interaction at this phase.
    The common blueshift of about $-1600\ {\rm km\ s^{-1}}$ shared by many emission lines from IMEs and IGEs irregardless of wavelength likely reflects a highly nonspherical geometry in the inner ejecta.
    }
    \label{fig:line_profiles}
\end{figure*}

The red shoulder gradually morphs into a second distinct emission feature at a velocity comparable to the blue excess, although it only becomes clearly discernible around $+290$ d. 
\cite{Michel_2025} ruled out the association of the red feature with [\ion{S}{2}] $\lambda6724$ or \ion{He}{1} $\lambda6678$ given its evolution with respect to the central component of H$\alpha$. 
Unlike the blue feature, the red one appears more flat-topped, extending roughly from $+4000\ {\rm km\ s^{-1}}$ to $+7000\ {\rm km\ s^{-1}}$. 
Both high-velocity features strengthen relative to the central emission and gradually decelerate toward the line center, eventually merging with the intermediate-width portion of the H$\alpha$ profile and settling near $\pm5000\ {\rm km\ s^{-1}}$ by $+749$ d. 
Similar evolution is shared by other H transitions, most notably Pa$\alpha$ and Br$\alpha$. 
We compare the line profiles of H$\alpha$, Pa$\alpha$ and Br$\alpha$ at three different epochs in the top row of Figure~\ref{fig:line_profiles}. 
The asymmetric H$\alpha$  profile also emerges in other H emission lines at around $+573$ d (e.g., H$\beta$ and H$\gamma$). 
By this late time, the bright central component of H$\alpha$ fades away, and H$\alpha$ becomes dominated by the persistent broad and asymmetric component between the two extremes at $\pm7000\ {\rm km\ s^{-1}}$.  
This transition in the H$\alpha$ profile coincides with the flattening of SN~2023ixf's optical light curves (see Figure~\ref{fig:lc}), marking the transition of the dominant heating source from radioactive decay to shock interaction (see Section~\ref{sec:shock_power}; \citealt{Jacobson-Galan_2025,Singh_2026}). 
Hereafter, we refer to these shock-powered, high-velocity emission features as ``CSM horns."

\begin{figure*}[t!]
    \centering
    \includegraphics[width=\textwidth]{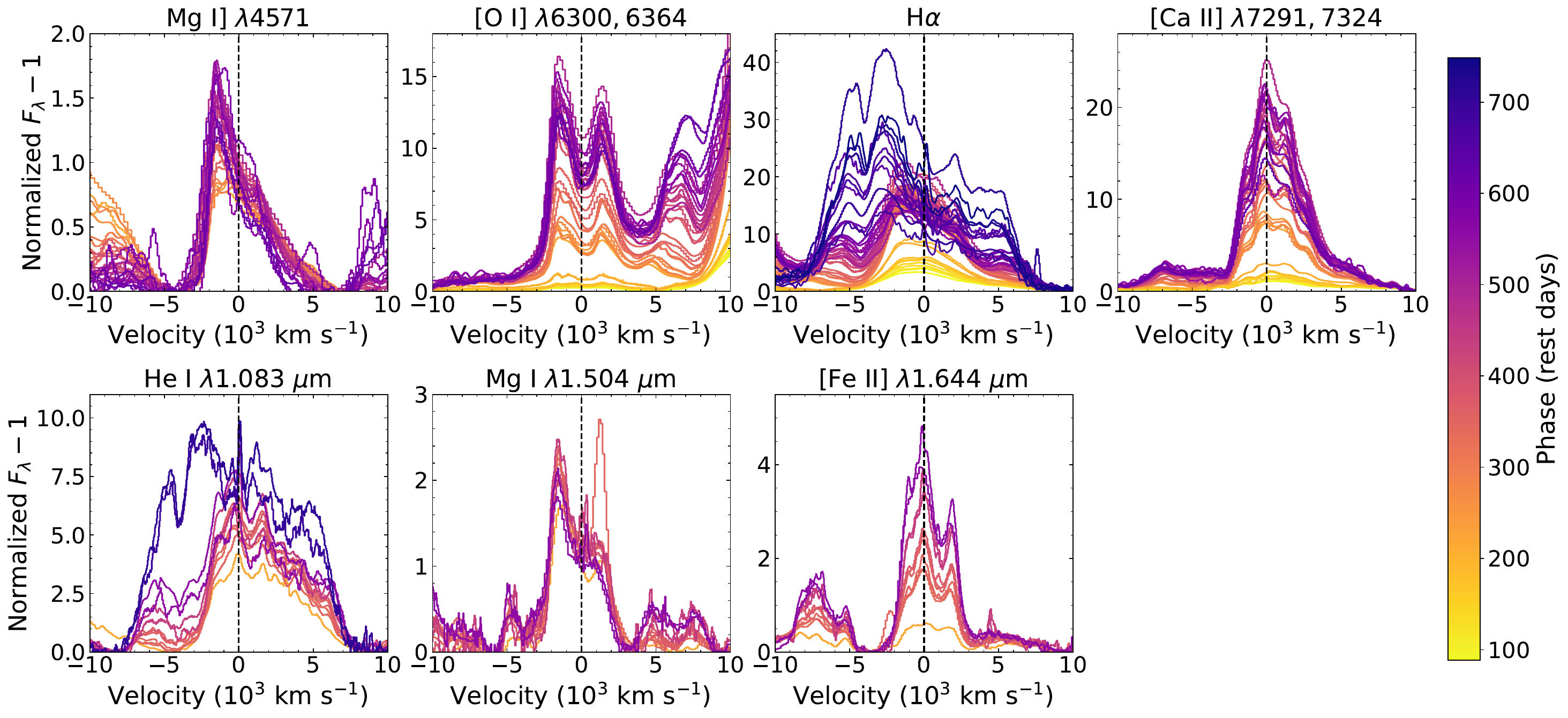}
    \caption{Temporal evolution of selected prominent emission lines in SN~2023ixf. 
    The rest-frame zero-velocity is marked with a black dashed line.
    The velocity of [\ion{O}{1}] $\lambda\lambda$6300, 6364 is measured with respect to the $\lambda$6300 line and the velocity of [\ion{Ca}{2}] $\lambda\lambda$7291, 7324 is measured with respect to the $\lambda$7291 line.
    Each line profile has been normalized to a linear pseudocontinuum.
    We note that \ion{He}{1} $1.083\ {\rm \mu m}$ is blended with Pa$\gamma$, and [\ion{Fe}{2}] $1.644\ {\rm \mu m}$ is likely heavily blended with emission from [\ion{Si}{2}] $1.645\ {\rm \mu m}$.
    The H$\alpha$ and \ion{He}{1} $1.083\ {\rm\mu m}$ profiles broaden over time, with high-velocity components emerging at around $+200$ d, consistent with the fast, outer ejecta interacting with low-density CSM. 
    In contrast, emission from O and Mg are persistently blueshifted, likely suggesting ejecta asymmetries or dust attenuation.
    }
    \label{fig:line_profiles_time}
\end{figure*}

Emission from \ion{He}{1} is also present in the nebular spectra of SN~2023ixf, with the strongest feature arising from \ion{He}{1} $1.083\ {\rm \mu m}$ that is blended with Pa$\gamma$ and [\ion{S}{1}] $1.082\ {\rm \mu m}$, as shown in Figure~\ref{fig:line_profiles_time}. 
Similar to H$\alpha$, a blue flux excess appears in \ion{He}{1} $1.083\ {\rm \mu m}$ at approximately $-6000\ {\rm km\ s^{-1}}$ at $+250$ d. 
A corresponding red excess is not clearly detected, likely due to blending with Pa$\gamma$. 
Over the next $\sim500$ days, the blue horn gradually decelerates and settles at approximately $-5000\ {\rm km\ s^{-1}}$ by $+695$ d. 
At late epochs ($+659$ d and $+695$ d), the line profile of \ion{He}{1} $1.083\ {\rm \mu m}$ closely resembles that of H$\alpha$ (see Figure~\ref{fig:line_profiles}), indicating that CSM interaction substantially contributes to the formation of He~{\sc i} emission as well. 
Evidence for CSM interaction in this line was first noted by \cite{Jacobson-Galan_2025}.

Additional, weaker emission lines of \ion{He}{1} $\lambda5876$, \ion{He}{1} $\lambda7065$, and \ion{He}{1} $2.058 {\rm \mu m}$ are also detected. For most of the nebular phase, \ion{He}{1} $\lambda5876$ is not clearly visible, as a significant fraction of its flux is expected to scatter into the \ion{Na}{1} D lines (\citealt{Jerkstrand_2015a, Ergon_2022}). 
At later epochs, however, a boxy and asymmetric line profile consistent with shock-powered \ion{He}{1} $\lambda5876$ becomes discernible, similar to the line profile observed in \ion{He}{1} 1.083 ${\rm \mu m}$. 
A similar argument applies to \ion{He}{1} $\lambda7065$, which is blended with the nearby [\ion{Ca}{2}] $\lambda\lambda7291$, 7324 doublet and initially shows negligible emission, but displays a clear blue peak near $-5000\ {\rm km\ s^{-1}}$ by $+573$ d. 
In contrast, the \ion{He}{1} $2.058\ {\rm \mu m}$ line exhibits a markedly different profile compared to other H and He lines. 
Instead of a blueshifted peak near $-2000\ {\rm km\ s^{-1}}$, \ion{He}{1} $2.058\ {\rm \mu m}$ displays a redshifted peak at $+2000\ {\rm km\ s^{-1}}$, with significant blue flux deficit. 
It is possible that an absorption component is also present, owing to the strong dependence of $2.058\ {\rm \mu m}$ on nonthermal excitation (\citealt{Li_McCray_1995,Li_2012,Jerkstrand_2015b}). 
At late epochs, \ion{He}{1} $2.058\ {\rm \mu m}$ emission also exhibits a high degree of fluctuations, suggesting that the He-emitting regions may possess a more complex geometry than implied by the other He features. 
We compare the profiles of these \ion{He}{1} lines to H$\alpha$ in the second row of Figure~\ref{fig:line_profiles}.

\subsection{Possible Carbon Emission}
\label{sec:carbon}

\begin{figure}[t!]
    \centering
    \includegraphics[width=\columnwidth]{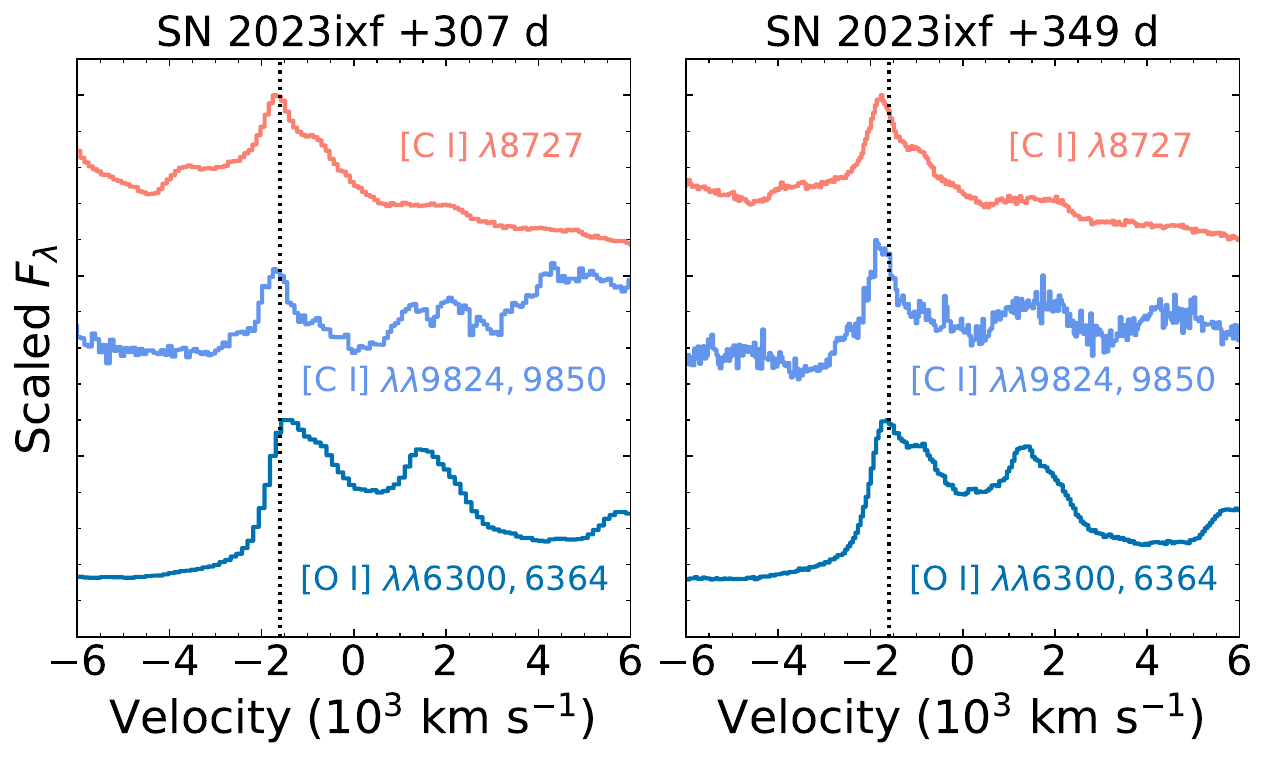}
    \caption{Line profiles of possible emission from [\ion{C}{1}] $\lambda8727$ and [\ion{C}{1}] $\lambda\lambda9824$, 9850 at $+307$ d and $+349$ d.
    These profiles are not normalized to the local pseudocontinua owing to blending with nearby emission (e.g., \ion{Ca}{2} NIR triplet in the case of [\ion{C}{1}] $\lambda8727$).
    The profile of [\ion{O}{1}] $\lambda\lambda $6300, 6364 is shown for comparison, which aligns with the blueshifted emission peak of [\ion{C}{1}] well, suggesting a common origin in the O/C layer.}
    \label{fig:CI_lines}
\end{figure}

We do not detect any strong emission from C in SN~2023ixf. However, a relatively weak emission peak that appears on the red side of the \ion{Ca}{2} NIR triplet can be attributed to [\ion{C}{1}] $\lambda8727$. 
This line has not been explicitly identified in previous studies of SN~2023ixf, possibly owing to differences in resolution and spectral coverage. 
Nevertheless, [\ion{C}{1}] $\lambda8727$ has been invoked in the past to explain the redshifted emission excess of the \ion{Ca}{2} NIR triplet in other CCSNe, including the Type Ic SN~2007gr (\citealt{Mazzali_2010}) and the samples of SNe~II analyzed by \cite{Maguire_2012} and \cite{Prentice_2022}, as well as in radiative-transfer models (\citealt{Jerkstrand_2015a, Ergon_2022, Dessart_2023b}). 
The identification of [\ion{C}{1}] $\lambda8727$ in SN~2023ixf is further supported by the presence of a feature consistent with the [\ion{C}{1}] $\lambda\lambda9824$, 9850 doublet in several of our optical spectra. 
The profiles of these two C emission lines, shown for two epochs in Figure~\ref{fig:CI_lines}, display similar kinematic properties. 
Both [\ion{C}{1}] $\lambda8727$ and [\ion{C}{1}] $\lambda\lambda9824$, 9850 exhibit asymmetric  profiles, with a blueshifted narrow core at $-1700\ {\rm km\ s^{-1}}$ superimposed on a broad base. In addition, a broad, redshifted emission component with a relatively flat top is present between $+500\ {\rm km\ s^{-1}}$ and $+3000\ {\rm km\ s^{-1}}$. 

As atomic cooling lines, both [\ion{C}{1}] $\lambda8727$ and [\ion{C}{1}] $\lambda\lambda9824$, 9850 are sensitive to physical conditions in the C-rich layers and, in particular, to the formation of CO (\citealt{Jerkstrand_2015a, Barmentloo_2026}). 
In the absence of CO, a substantial fraction of cooling occurs through [\ion{C}{1}], resulting in stronger emission. 
In SN~2023ixf, however, CO emission was detected as early as $+199$ d (\citealt{Park_2025}), indicating that molecule formation has already commenced. 
The continued detection of [\ion{C}{1}] emission at later epochs may therefore reflect spatially distinct emitting zones for C and CO. 
We additionally include the line profile of [\ion{O}{1}] $\lambda\lambda $6300, 6364 in Figure~\ref{fig:CI_lines} to compare the emitting regions of C and O. 
The emission peaks of [\ion{C}{1}] line up well with [\ion{O}{1}], perhaps suggesting a common origin exterior to the CO formation site. 
Recent work by \cite{Barmentloo_2026} has also explored the use of [\ion{C}{1}] $\lambda8727$ and [\ion{C}{1}] $\lambda\lambda9824$, 9850 as diagnostics of carbon core mass in stripped-envelope SNe (SESNe). 

\subsection{IMEs: Revealing the Inner Ejecta}
\label{sec:oxygen_magnesium}

The nebular-phase spectra of SN~2023ixf exhibit prominent emission from IMEs such as O, Mg, Ne, Ar, and Ca, which together trace the composition and structure of the inner ejecta. 
Emission from [\ion{O}{1}] $\lambda\lambda$6300, 6364 is one of the strongest features in the optical spectra of SN~2023ixf, which emerges almost immediately after the drop from the plateau. 
Between $+89$ d and $+113$ d, the [\ion{O}{1}] doublet appears largely blended, although two individual peaks symmetric about 6300 \AA\ are distinguishable. By $+138$ d, a clear double-peaked structure separated by $\sim 64\ {\rm\AA}$ (corresponding with the wavelength separation of the [\ion{O}{1}] $\lambda6300$ and $\lambda6364$ lines) appears, with a blueshift of $\sim 1600\ {\rm km\ s^{-1}}$ that persists throughout the entire nebular phase. 
The blueshifted [\ion{O}{1}] $\lambda6300$ line itself has a distinct ``notch", which is likely due to telluric absorption by O$_2$ based on several of our higher-resolution MMT/Blue Channel spectra and the echelle spectra presented by \cite{Smith_2023}. 
A short-lived, boxy emission at $\sim 6150\ {\rm \AA}$ that fades within the first 50 days of the nebular phase can be attributed to \ion{Ba}{2} $\lambda 6142$. 
The early appearance of [\ion{O}{1}] may also point to a partially stripped progenitor (\citealt{Elmhamdi_2011}) for SN~2023ixf, which is consistent with previous studies of light-curve and radiative-transfer modeling (e.g., \citealt{Fang_2025b,Hsu_2025,Dessart_2026_photo}). 
Additional emission lines of O, both forbidden and permitted, are also detected, including [\ion{O}{1}] $\lambda5577$, \ion{O}{1} $\lambda7774$, \ion{O}{1} $\lambda8446$, \ion{O}{1} $\lambda9264$, and \ion{O}{1} 1.129 $\mu{\rm m}$. 
However, many of these identifications are subject to significant uncertainties owing to blending and line shifts. For example, the blueshifted emission near [\ion{O}{1}] $\lambda5577$ may instead arise from [\ion{Fe}{2}] or [\ion{Co}{2}] (\citealt{Houck_1996,Milisavljevic_2010, Jerkstrand_2014}), \ion{O}{1} $\lambda7774$ could be contaminated by telluric absorptions and \ion{K}{1} $\lambda\lambda7665$, 7699 (\citealt{Chornock_2010,Dessart_2013,Silverman_2017}), and \ion{O}{1} 1.129 $\mu{\rm m}$ is likely blended with [\ion{S}{1}] 1.131 $\mu{\rm m}$. 
Most of these weaker O emission lines fade or become very weak by $\sim+300$ d.

Emission from Mg becomes prominent later in the nebular phase compared to O. 
Beginning at $+199$ d, \ion{Mg}{1}] $\lambda4571$ emerges as a blueshifted feature with a slanted profile and an extended red tail. 
The \ion{Mg}{1}] $\lambda4571$ flux relative to the local continuum increases as the ejecta expand. 
At these early epochs, however, the emission may still be partially blended with lines from IGEs (\citealt{Maeda_2006,Jerkstrand_2015a}). 
The line profile of \ion{Mg}{1}] $\lambda4571$ decreases in width between $+248$ d and $+578$ d, before fading by $\sim +600$~d. 
In the NIR, strong recombination emission from \ion{Mg}{1} 1.504 $\mu{\rm m}$ is detected at $+200$ d, although its earliest emergence can be traced back to $+81$ d post-explosion (\citealt{Li_2025}). Both \ion{Mg}{1}] $\lambda4571$ and \ion{Mg}{1} 1.504 $\mu{\rm m}$ exhibit persistent blueshifts of approximately $-1600\ {\rm km\ s^{-1}}$, consistent with the offset observed in [\ion{O}{1}] $\lambda\lambda$6300, 6364 and suggesting a common origin within the O-rich ejecta. 
A clear redshifted peak at $+1600\ {\rm km\ s^{-1}}$ is also visible in the \ion{Mg}{1} 1.504 $\mu{\rm m}$ line in the $+200$ d spectrum, which could indicate a bipolar distribution of material or inhomogeneous heating. 
Several weaker Mg features are also tentatively identified, including \ion{Mg}{1} 1.488, 1.577, 1.711 $\mu$m in the ground-based NIR spectra, and \ion{Mg}{1} 2.486, 3.867, 4.200 $\mu$m in the $+252$ d and $+375$ d JWST spectra, the latter three of which were not previously identified. 
At late times when the heating source is dominated by CSM interaction, we also see the emergence of \ion{Mg}{2} $\lambda\lambda 9218, 9244$, which likely formed in the swept-up cold dense shell (CDS; \citealt{Dessart_2023,Bostroem_2025, Jacobson-Galan_2025}).

One strong Ne emission line is detected, [\ion{Ne}{2}] $12.810\ {\rm \mu m}$ \citep{Medler_2025}. 
Radiative-transfer models \citep{Dessart_2025,Dessart_2025b} predict this emission to be among the strongest MIR features, originating predominantly from $^{20}$Ne in the O/Ne/Mg shell. 
While the measured full width at half-maximum intensity (FWHM) is broadly consistent with those of [\ion{O}{1}] $\lambda\lambda$6300, 6364, \ion{Mg}{1}] $\lambda4571$, and \ion{Mg}{1} 1.504 $\mu{\rm m}$, [\ion{Ne}{2}] $12.810\ {\rm \mu m}$ has a centroid and peak at roughly zero velocity, rather than blueshifted like other IME lines (\citealt{Medler_2025}). 
This discrepancy may reflect the limited spectral resolution of MIRI/LRS, geometric effects within the ejecta, or ionization stratification, as discussed in detail by \cite{Dessart_2025b}. 
Several Na features are detected in SN~2023ixf, most notably the \ion{Na}{1} $\lambda\lambda 5890,5896$ doublet possibly blended with \ion{He}{1} $\lambda 5876$, which has a strong P~Cygni component and a double-peaked profile (\citealt{Li_2025}) for most of the nebular phase before transforming into boxy \ion{He}{1} $\lambda 5876$ emission at $\sim +600$ d. 
In the NIR, weak and boxy emission from \ion{Na}{1} $2.206\ {\rm \mu m}$ blends with the red wing of Br$\gamma$.

Similar to the [\ion{O}{1}] doublet, the [\ion{Ca}{2}] $\lambda\lambda7281, 7324$ doublet and the \ion{Ca}{2} $\lambda\lambda8498$, 8542, 8662 NIR triplet are visible at the start of the nebular phase. 
Based on optical spectra during the photospheric phase \citep{Singh_2024}, however, the [\ion{Ca}{2}] doublet was detected as early as $+81$ d and the \ion{Ca}{2} triplet as early as $+26$ d. 
The intensity of the forbidden [\ion{Ca}{2}] doublet grows stronger relative to the permitted \ion{Ca}{2} NIR triplet as the electron density drops in the ejecta. 
Both [\ion{Ca}{2}] and \ion{Ca}{2} appear to have peaks near zero velocity throughout the nebular phase, contrary to the $-1600\ {\rm km\ s^{-1}}$ blueshift shared by other prominent emission features (see Figure~\ref{fig:line_profiles}). 
The region around 7921 \AA\ is dominated by the [\ion{Ca}{2}] doublet, with noticeable excesses on both wings, which have been associated with a series of weak [\ion{Fe}{2}] lines and [\ion{Ni}{2}] emission from stable $^{58}$Ni in the past (\citealt{Jerkstrand_2015b, Terreran_2016, Michel_2025}). 
There is also a blue shoulder around $7260\ {\rm \AA}$ that is consistent with \ion{He}{1} $\lambda7281$ offset by $-900\ {\rm km\ s^{-1}}$, although this is low compared to the blueshift of $-1600\ {\rm km\ s^{-1}}$ for the \ion{He}{1} $\lambda 7065$ emission nearby. 
If this line is indeed \ion{He}{1} $\lambda7281$, then it is also observed to be much stronger than theoretical predictions (\citealt{Jerkstrand_2015b}). 
Therefore, it is more likely that this extra emission at $\sim 7260\ {\rm \AA}$ is a blueshifted component of the [\ion{Ca}{2}] $\lambda 7291$ emission.

While \citet{Dessart_2025b} predicts negligible emission from Si, other works have identified various Si emission lines, mostly in the NIR (e.g., \citealt{Jerkstrand_2015b, Dessart_Hillier_2020b}), albeit heavily blended with other features. 
Most notably, we detect emission from [\ion{Si}{1}] $1.606\ {\rm \mu m}$ and [\ion{Si}{1}] $1.645\ {\rm \mu m}$ blended with nearby [\ion{Fe}{2}] emission features (see Section~\ref{sec:iges}), which has been proposed in the past to explain the emission complex at $\sim 1.6-1.65\ {\rm \mu m}$ (\citealt{Oliva_1987}). 
This is in part reinforced by the detection of \ion{Si}{1} $1.203\ {\rm \mu m}$. 
Lastly, other IMEs contribute noticeable emission through [\ion{S}{1}] 1.082, 1.131\ ${\rm\mu m}$ (both are blended with other features) and unresolved [\ion{Ar}{2}] $6.983\ {\rm \mu m}$.

\subsection{IGEs: Clues to Explosion Geometry and Mixing}
\label{sec:iges}

Radiative-transfer models of SNe II predict a forest of permitted and forbidden iron lines, primarily from \ion{Fe}{1} and \ion{Fe}{2}. 
In the optical, no isolated Fe line is detected. 
In the NIR, emission from [\ion{Fe}{2}] $1.257\ {\rm \mu m}$ and [\ion{Fe}{2}] $1.644\ {\rm \mu m}$ are the strongest features, where [\ion{Fe}{2}] $1.257\ {\rm \mu m}$ is blended with the red shoulder of Pa$\beta$. 
However, the observed intensity of [\ion{Fe}{2}] $1.644\ {\rm \mu m}$ is much stronger than that of [\ion{Fe}{2}] $1.257\ {\rm \mu m}$, which is in tension with the intensity ratio predicted by theory and previous observations ($I_{1.644\ {\rm \mu m}}/I_{1.257\ {\rm \mu m}}\approx 0.7-0.77$; \citealt{Oliva_1987,Nussbaumer_Storey_1988,sh06}). 
In addition, the emission complex around 1.6--$1.65\ {\rm \mu m}$ consists of two nearly identical line profiles consistent with [\ion{Si}{1}] 1.606, 1.645 ${\rm\mu m}$. 
We therefore conclude that [\ion{Si}{1}] $1.645\ {\rm\mu m}$ makes a nonnegligible contribution to the emission structure at $\sim 1.66-1.67\ {\rm\mu m}$, contrary to model predictions (\citealt{Dessart_2025b}). 
A similar argument likely applies to the multipeaked [\ion{Fe}{2}] line profiles recently reported in the nebular spectra of SN~2024ggi, given the mismatch in velocity between [\ion{Fe}{2}] $1.257\ {\rm \mu m}$ and [\ion{Fe}{2}] $1.644\ {\rm \mu m}$ (\citealt{Ferrari_2025,Hueichapan_2025}). 
In addition, we detect [\ion{Fe}{2}] $1.321\ {\rm \mu m}$, [\ion{Fe}{2}] $1.534\ {\rm \mu m}$, [\ion{Fe}{2}] $1.599\ {\rm \mu m}$, \ion{Fe}{2} $1.677\ {\rm \mu m}$, and \ion{Fe}{2} $1.809\ {\rm \mu m}$. 

During the nebular phase, all of the radioactive $^{56}$Ni synthesized during the explosion has decayed (with a half-life of 6.1 days; \citealt{Nadyozhin_1994}), so any Ni emission must originate only from stable isotopes like $^{58}$Ni and $^{60}$Ni. 
In SN~2023ixf, \cite{Medler_2025} identified strong emission from [\ion{Ni}{1}] $3.12\ {\rm \mu m}$, [\ion{Ni}{1}] $11.308\ {\rm \mu m}$, [\ion{Ni}{2}] $1.939\ {\rm \mu m}$, and [\ion{Ni}{2}] $6.636\ {\rm \mu m}$. 
Similar to its optical counterparts (e.g., [\ion{Ni}{2}] $\lambda\lambda 7378$, 7412), [\ion{Ni}{2}] 1.939  ${\rm \mu m}$ is blended with the red wing of Pa$\alpha$ and [\ion{Ni}{2}] $6.636\ {\rm \mu m}$ is unresolved owing to the resolution of MIRI/LRS. 
We note that the emission from Br$\delta$ has likely subsided in the $+374$ d spectrum, given the diminishing flux in both Br$\beta$ and Br$\gamma$. 
Therefore, we suspect the emission structure around  1.92--1.95 ${\rm \mu m}$ is dominated by [\ion{Ni}{2}] $1.939\ {\rm \mu m}$ at $+374$ d.
Lastly, strong emission from [\ion{Co}{2}] $10.521\ {\rm \mu m}$ is blended with the nearby [\ion{Ni}{2}] $10.682\ {\rm \mu m}$ is detected in the $+252$ d and $+374$ d JWST spectra.

\begin{figure}[t!]
    \centering
    \includegraphics[width=\columnwidth]{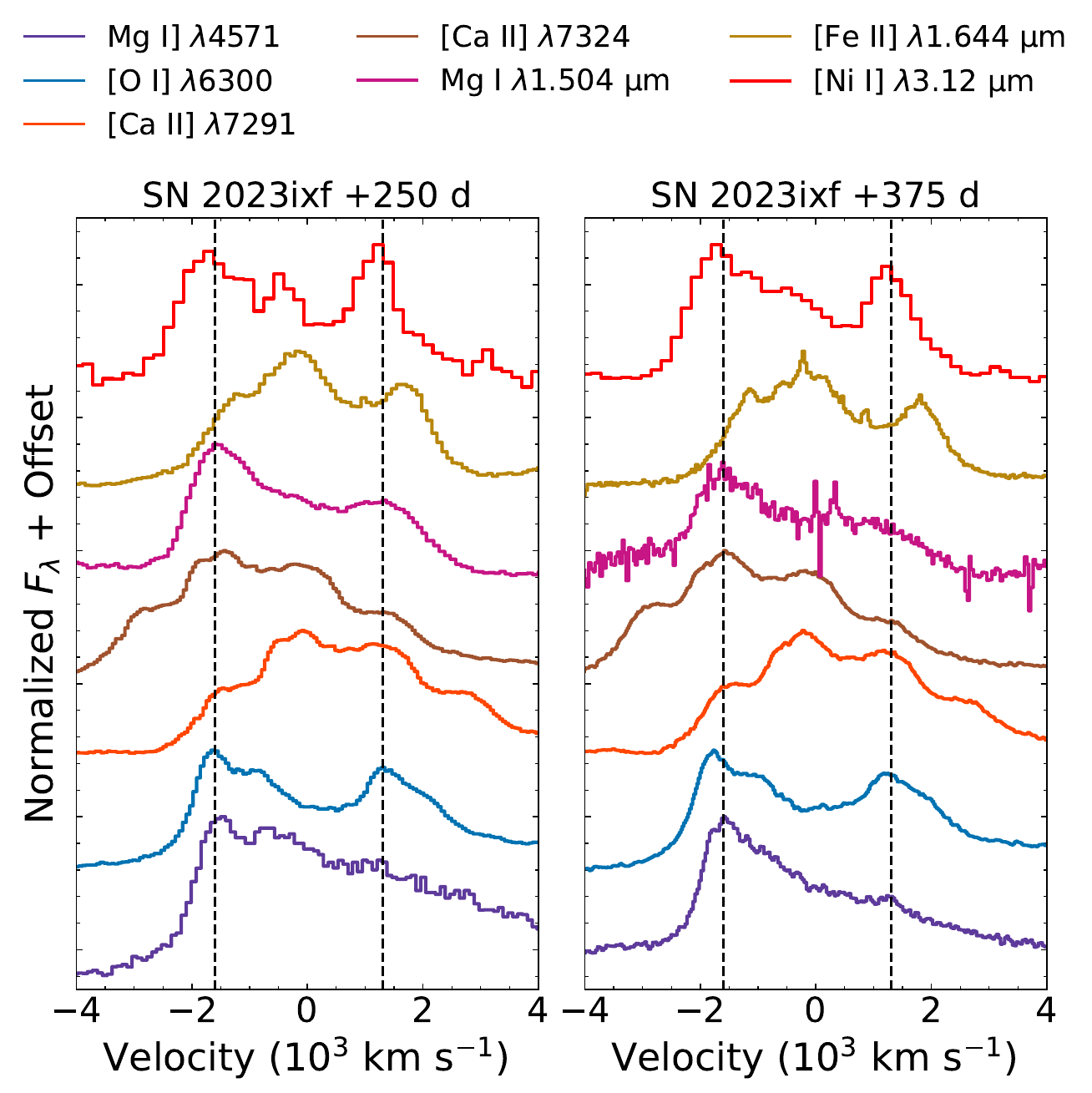}
    \caption{Line-profile comparison of several features in SN~2023ixf at around $+250$ d and $+375$ d after explosion. 
    The optical and NIR spectra used in each panel are within $\pm7$ days from each other.
    The vertical dashed lines denote the two emission peaks of [\ion{Ni}{1}] $3.12\ {\rm \mu m}$ at $-1500\ {\rm km\ s^{-1}}$ and $+1300\ {\rm km\ s^{-1}}$. 
    Given that the two peaks of [\ion{Ni}{1}] $3.12\ {\rm \mu m}$ line up well with various emission lines from IMEs, it is possible that SN~2023ixf has an asymmetric distribution of Ni that leads to inhomogeneous heating of the ejecta.}
    \label{fig:multipeaks}
\end{figure}

Originally, \cite{Medler_2025} identified the emission complex around 3.05--3.20 ${\rm \mu m}$ as a blend of [\ion{Ni}{1}] $3.12\ {\rm \mu m}$ with [\ion{S}{1}] $3.107\ {\rm \mu m}$ and [\ion{Ar}{2}] $3.137\ {\rm \mu m}$. 
However, the FWHM of the blueshifted component that corresponds to [\ion{Ar}{2}] $3.137\ {\rm \mu m}$ is much narrower than that of [\ion{Ar}{2}] $6.985\ {\rm \mu m}$ (\citealt{Medler_2025}), and these lines are not predicted to be strong in theoretical models of \cite{Dessart_2025b} and \cite{Dessart_2025c}. 
An alternative explanation for the double-peaked emission around [\ion{Ni}{1}] $3.12\ {\rm \mu m}$ is bipolar ejection or clumping of IGEs due to the nickel-bubble effect (\citealt{Woosley_1988,Basko_1994,Kifonidis_2000,Wang_2005, Dessart_2018}). 
A doubled-peaked [\ion{Ni}{1}] $3.12\ {\rm \mu m}$ line profile was also reported in SN~2024ggi (\citealt{Dessart_2025,Jacobson-Galan_2026}). 
An asymmetric distribution of Ni could also lead to inhomogeneous heating of the ejecta and multipeaked line profiles in other elements (\citealt{Gerardy_2000,Utrobin_Chugai_2017, Bose_2019,Utrobin_2021}) and partially explain late-time spectropolarimetric observations of SN~2023ixf (\citealt{Singh_2024, Shrestha_2025, Vasylyev_2026}). 
We compare the line profile of [\ion{Ni}{1}] $3.12\ {\rm \mu m}$ to selected emission lines in Figure~\ref{fig:multipeaks}. 
Qualitatively, the two distinct peaks of [\ion{Ni}{1}] $3.12\ {\rm \mu m}$ at $-1600\ {\rm km\ s^{-1}}$ and $+1300\ {\rm km\ s^{-1}}$ line up reasonably well with the emission peaks of several IMEs, supporting an intrinsically asymmetric ejecta. 

\section{The H$\alpha$ Evolution}
\label{sec:Halpha}

\subsection{From Decay-powered to Shock-powered Emission}
\label{sec:shock_power}

The CSM horns in various H and He lines likely arise from X-rays generated by the reverse shock, with half of the energy thermalized by the CDS (\citealt{Fransson_1996}) and the other half ionizing the fast, unshocked ejecta (\citealt{Chevalier_Fransson_1994, Smith_2008}). 
Here, we aim to trace the evolution of the shock power in SN~2023ixf using our extensive spectral time series.

To calculate the shock power $L_{\rm sh}$, we use Eq.~(1) of \cite{Bostroem_2025}, which is based on the analytic formalism of \cite{Fransson_1996}, assuming a steady-state wind density profile with $s=2$ and a power-law density profile with $n=12$ for a RSG envelope  (\citealt{Chevalier_Fransson_2017}).
For simplicity, we assume a constant wind velocity of $v_{\rm wind}=25\ {\rm km\ s^{-1}}$ measured from early high-resolution spectra (\citealt{Dickinson_2025}) and a constant wind mass-loss rate of $\dot{M}\approx 10^{-4}\ M_{\odot}\ {\rm yr^{-1}}$ based on X-ray observations (\citealt{Nayana_2025, Jacobson-Galan_2025}).
We note that this is an extremely high mass-loss rate compared to normal RSGs \citep{beasor20}.
The calculation of the ejecta velocity at the reverse shock is not as straightforward.
\cite{Bostroem_2025} used the blue velocity limit of $V_{\rm ej}\approx (7000-10,000)\ {\rm km\ s^{-1}}$ from the \ion{Mg}{2} $\lambda\lambda 2796$, 2803 doublet that corresponded to 99\% of the integrated flux as a proxy for the reverse-shock velocity.
On the other hand, \cite{Jacobson-Galan_2025} assumed $V_{\rm ej}=6500\ {\rm km\ s^{-1}}$ with an additional $1000\ {\rm km\ s^{-1}}$ in uncertainty to cover the decelerating boxy H$\alpha$ emission.
Here, we take an intermediate approach by measuring the velocity of the blue emission peak $V_{\rm peak, blue}$, as well as the blue velocity edge $V_{\rm max, blue}$ that encompasses 99\% of the integrated H$\alpha$ flux as done by \cite{Bostroem_2025}, which results in $\left|V_{\rm peak, blue}\right|<\left|V_{\rm max, blue}\right|$ at all epochs.
We then perform a simple Monte Carlo simulation assuming $V_{\rm ej}\approx U(\left|V_{\rm peak, blue}\right|, \left|V_{\rm max, blue}\right|)$, calculating $L_{\rm sh}$ for each randomly sampled $V_{\rm ej}$, and repeat this process 10,000 times for each epoch.

\begin{figure}[t!]
    \centering
    \includegraphics[width=\columnwidth]{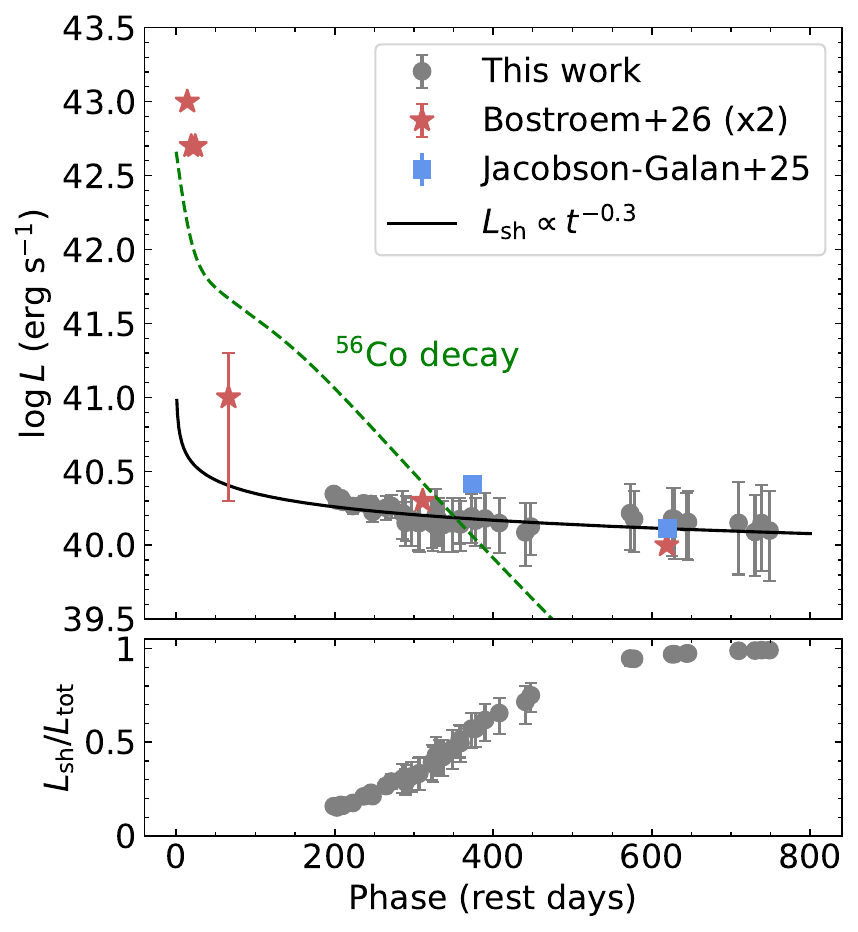}
    \caption{{\it Top:} The evolution of shock-power luminosity in SN~2023ixf (gray circles) derived from the H$\alpha$ velocity, assuming a constant mass-loss rate of $\dot{M}=10^{-4}\ M_{\odot}\ {\rm yr^{-1}}$ and a wind velocity of $v_{\rm wind}=25\ {\rm km\ s^{-1}}$.
    The error bars correspond to $1\sigma$ uncertainty bounds calculated from a Monte Carlo procedure with 10,000 samples to estimate the ejecta velocity at the reverse shock.
    Estimates from UV observations (red stars; \citealt{Bostroem_2025}) and UV-through-NIR spectral energy distributions (blue squares; \citealt{Jacobson-Galan_2025}) are shown for comparison.
    Note that a factor of 2 has been applied to the $L_{\rm sh}$ estimates from \cite{Bostroem_2025}, as their derived results only account for half of the reverse-shock luminosity.
    We also show the expected $\gamma$-ray deposition rate from $^{56}$Co decay in dashed green using $M_{\rm Ni}=0.058\ M_{\odot}$ and $T_0=264.6$ days from \cite{Jacobson-Galan_2025}.
    The late-time shock luminosity in SN~2023ixf is consistent with a radiative reverse shock.
    {\it Bottom:} The corresponding fraction of shock power relative to the total power in SN~2023ixf, where $L_{\rm tot}=L_{\rm sh}+L_{\rm decay}$.
    By $+600$ d, most of the luminosity in SN~2023ixf comes from CSM interaction.}
    \label{fig:shock_power}
\end{figure}

The resulting evolution of $L_{\rm sh}$ (median and $1\sigma$ bounds) is shown in Figure~\ref{fig:shock_power}, along with estimates independently derived by \cite{Bostroem_2025} and \cite{Jacobson-Galan_2025}.
Overall, our calculations trace the $L_{\rm sh}\propto t^{-0.3}$ evolution expected for a radiative reverse shock (\citealt{Fransson_1996}) and are consistent with the results from \cite{Bostroem_2025} and \cite{Jacobson-Galan_2025}. 
To demonstrate the transition from decay to shock power, we overplot the expected $^{56}$Co decay luminosity and the fraction of $L_{\rm sh}$ as a function of time in Figure~\ref{fig:shock_power}.
The decay luminosity is calculated following the formalism of \cite{Wheeler_2015} and assuming a $^{56}$Ni mass of $M_{\rm Ni}=0.058\ M_{\odot}$ and $T_0 = 264.6$ days from \cite{Jacobson-Galan_2025}.
Coincident with the flattening of optical light curves, we find that by around $+600$ d, almost all of the heating comes from shock power.

\subsection{Distinct Line-profile Components?}
\label{sec:H_profiles}

\begin{figure*}[t!]
    \centering
    \includegraphics[width=.9\textwidth]{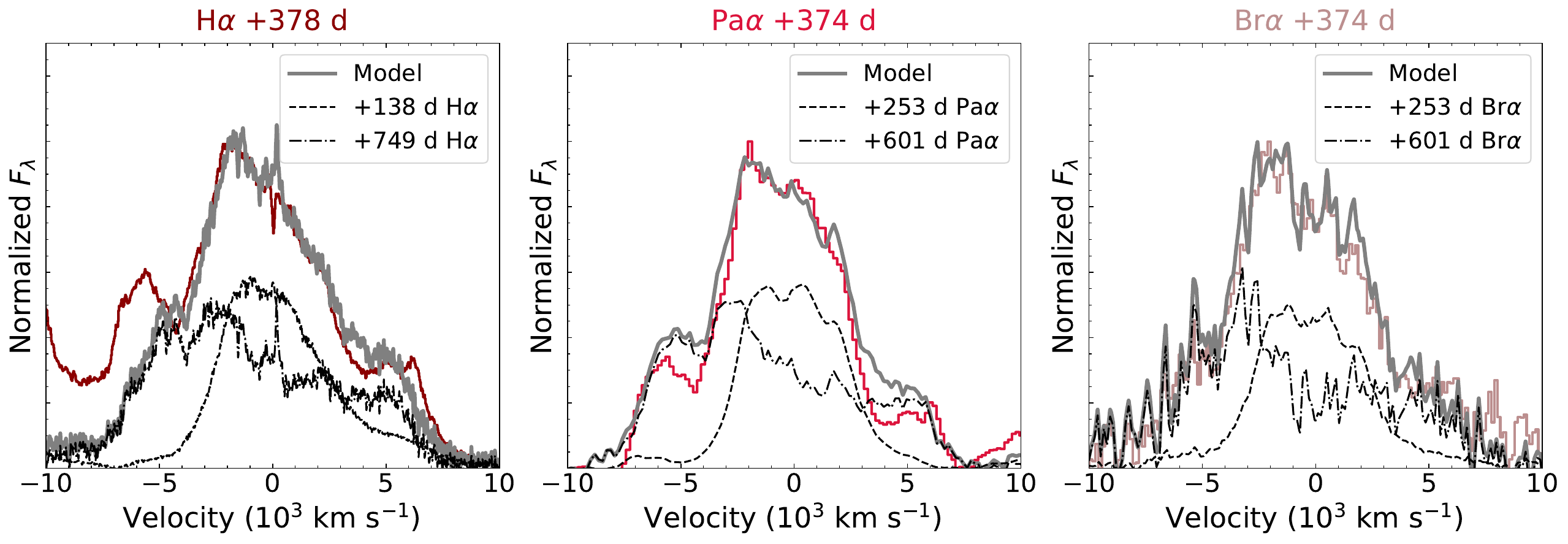}
    \caption{Line profiles of H$\alpha$, Pa$\alpha$, and Br$\alpha$ around $+375$ d. 
    The profiles at this intermediate epoch (where both decay-powered emission and  shock-powered emission are present) can be fit by scaling and combining line profiles from an earlier epoch with mostly decay-powered emission and a later epoch with mostly shock-powered emission.
    Through this simple scaling exercise, we find that the emission excess at $\approx-3000\ {\rm km\ s^{-1}}$ in the $+378$ H$\alpha$ profile can be naturally explained by the intermediate-width emission peak at late times, once the decay luminosity from $^{56}$Co have faded substantially.
    The observed H$\alpha$ evolution is similar to those of SN~1998S (\citealt{Gerardy_2000, Pozzo_2004}) and PTF11iqb (\citealt{Smith_2015}), which likely arises from the dense, aspherical CSM around SN~2023ixf's progenitor that has been swept-up by the forward shock.
    }
    \label{fig:H_scaled}
\end{figure*}

Several previous studies of SN~2023ixf have already identified distinct emission peaks in multiple H emission lines --- most notably, the triple-peaked central emission from the ejecta powered by $^{56}$Co decay (\citealt{Singh_2024, Kumar_2025, Medler_2025}).
By $+378$ d, we see an additional blue flux excess appears at $\sim -3000\ {\rm km\ s^{-1}}$ (distinct from the high-velocity CSM horn at $\sim -6000\ {\rm km\ s^{-1}}$ at this epoch), which could be associated with broad emission from [\ion{N}{2}] $\lambda\lambda6548$, 6583 in the H envelope or blanketing effects (\citealt{Hoeflich_1988}).
However, at this epoch, [\ion{N}{2}] emission is expected to be weak compared to H$\alpha$, and \cite{Medler_2025} argued against line blanketing as the root cause owing to the presence of identical peaks in other sufficiently isolated H emission lines. 
Furthermore, the blue excess at $\sim -3000\ {\rm km\ s^{-1}}$ at $+378$ d lines up well with the H$\alpha$ emission peak at later epochs ($\gtrsim +600$ d, after the light curves have flattened), when shock-powered emission dominates, and decay-powered nebular emission from the ejecta has largely subsided. 

We hypothesize that this ``extra" emission is part of the late-time shocked-powered emission structure ($-4000\ {\rm km\ s^{-1}}\lesssim v\lesssim 4000\ {\rm km\ s^{-1}}$) that emerges after $+573$ d in our spectral sequence, which is weaker compared to the earlier decay-powered emission.
We test our hypothesis by scaling and summing our higher-resolution H$\alpha$ line profiles at $+138$ d ($\approx90\%$ decay-powered) and at $+749$ d ($\approx 99\%$ shock-powered), where the ratio is determined by scaling the decay and shock power at $+378$ d. 
Qualitatively, the asymmetric H$\alpha$ profile at $+378$ d is well matched by a linear combination of the H$\alpha$ profiles at $+139$ d and $+749$ d.
We find similarly reasonable agreements for the $+374$ Pa$\alpha$ and Br$\alpha$ profiles modeled in the same fashion.
We show these fits in Figure~\ref{fig:H_scaled}.

The results from our simple scaling exercise reveal two key insights.
First, many asymmetric H line profiles could be decomposed into three primary components: the central core powered by radioactivity, high-velocity CSM horns powered by the shock, and an intermediate-width component that progressively reveals itself as the decay power fades.
The overall H$\alpha$ evolution is similar to that of the Type IIn SN~1998S (\citealt{Gerardy_2000,Pozzo_2004}) and the transitional SN II PTF11iqb (albeit with a persistently blueshifted peak; \citealt{Smith_2015}).  
To that end, the asymmetric late-time H$\alpha$ profile is consistent with being generated by an aspherical shock front as suggested by previous studies (\citealt{Ferrari_2024,Singh_2024,Fang_2025b,Kumar_2025}), with a faster shock front ahead of the freely expanding ejecta and a slower shock front decelerated by an initially aspherical dense CSM (e.g., an equatorial disk or torus).
Furthermore, given the lack of narrow emission lines in late-time observations, it is likely that the dense CSM has already been swept-up by the forward shock.
Several other lines of evidence also agree with an aspherical CSM for SN~2023ixf, such as early narrow emission features (\citealt{Smith_2023, Dickinson_2025}) and spectropolarimetric signals (\citealt{Vasylyev_2023, Vasylyev_2026, Singh_2024,Shrestha_2025}).
Second, we find that the slanted profiles of H$\alpha$, Pa$\alpha$, and Br$\alpha$ around $+375$ d can be  reproduced by carefully scaling the decay-powered and shock-powered emission components without the need for invoking dust attenuation.
It is, however,  possible that dust forms early in SN~2023ixf and the decay-powered H$\alpha$ profile at $+138$ has already been attenuated.
We discuss our interpretation for the CSM configuration and the effects of dust in more detail in Section~\ref{sec:discussion}.

\section{Line-profile Analysis}
\label{sec:line_profile_analysis}

The opacity drops significantly during the nebular phase, and emission-line profiles reveal the geometry and composition of the ejecta. 
In the following section, we follow the line-profile fitting procedure outlined by \cite{Kwok_2023} (using {\tt scipy.optimize.curve\_fit}; \citealt{Virtanen_2020}) to fit the profiles of a few prominent or interesting emission features, motivated primarily by geometric configurations. 
If multiple lines from a particular atom or ion fall within the fitted wavelength region of interest (for example, the [\ion{O}{1}] $\lambda\lambda$6300, 6364 doublet), they are required to have the same parameters, including offset, FWHM, and profile shape. 
As the main goal of this work is to constrain the ejecta geometry and elemental distribution, we focus the bulk of our analysis on observations around $+250$ d and $+375$ d where radioactive decay still dominates the heating source and where full optical-to-MIR coverage is available.
Note that while several studies have implicated the presence of dust in the ejecta of SN~2023ixf through CO emission, strong thermal infrared (IR) excess, and time-dependent asymmetries in emission-line profiles (\citealt{Jacobson-Galan_2025, Medler_2025, Park_2025, Singh_2026}), we do not model the effects of dust absorption or scattering in this section, as detailed radiative transfer is required. 
We discuss how dust formation affects the derived ejecta structure in Section~\ref{sec:dust}.

\subsection{Double-peaked IME and IGE Emission}
\label{sec:double_peaked}

\begin{figure*}[t!]
    \centering
    \includegraphics[width=.9\textwidth]{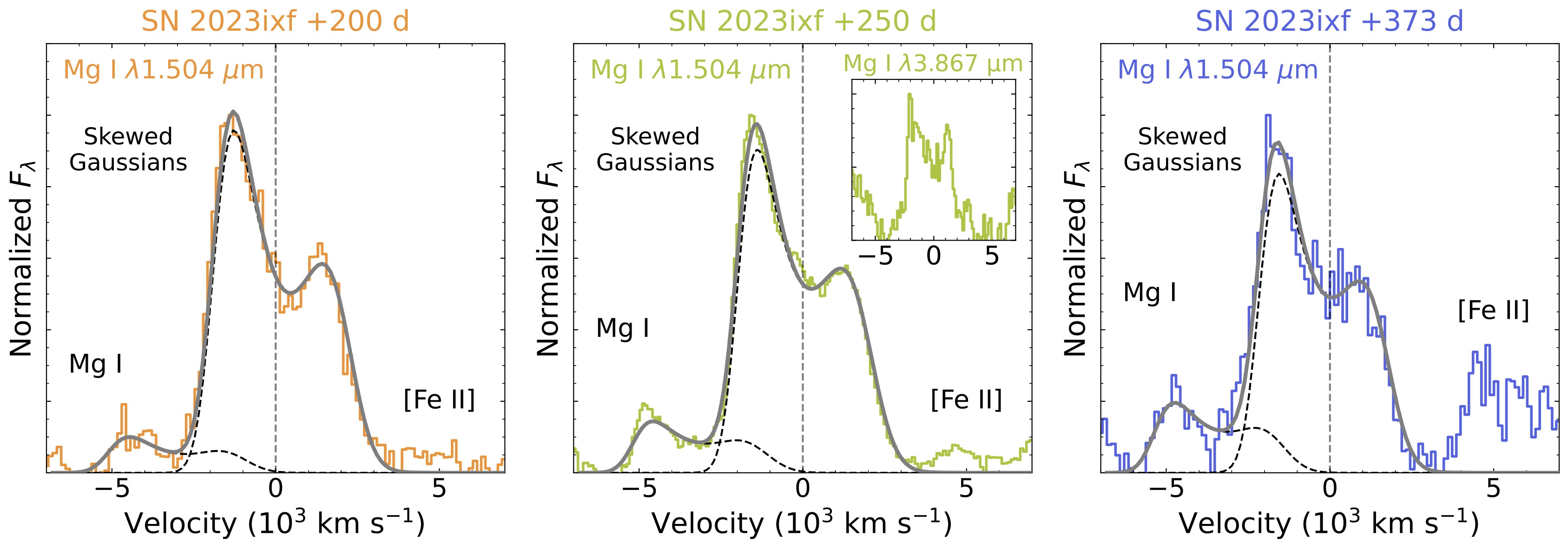}
    \includegraphics[width=.9\textwidth]{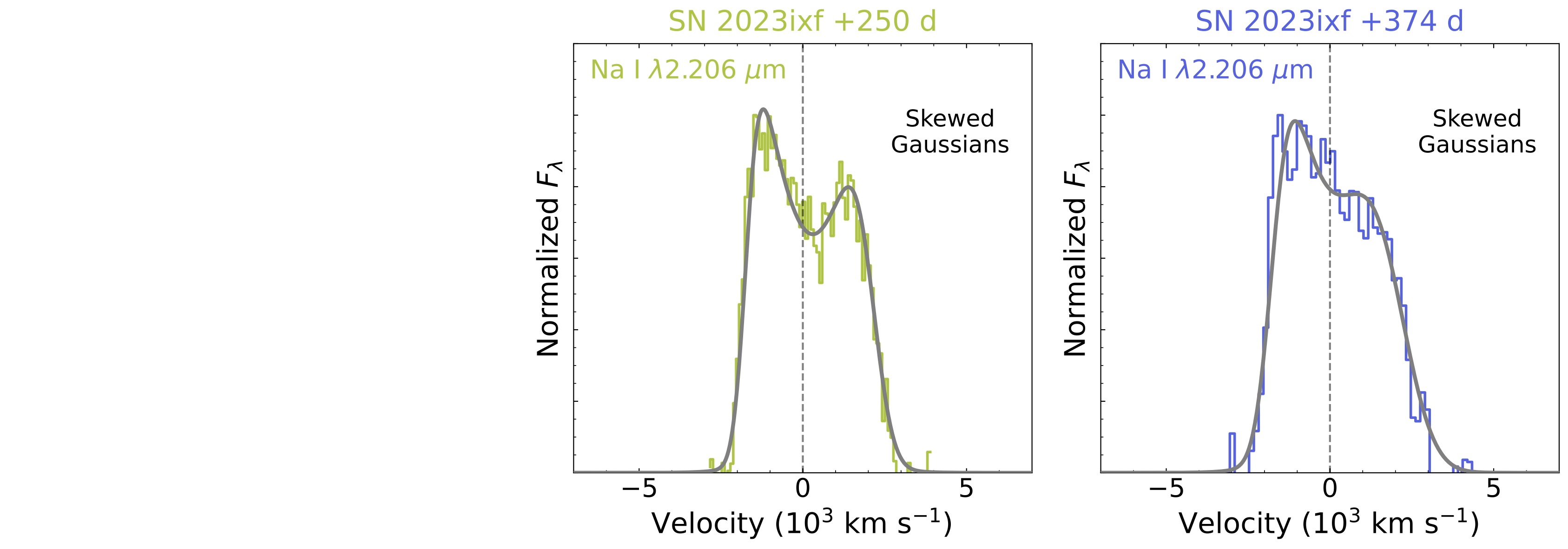}
    \includegraphics[width=.9\textwidth]{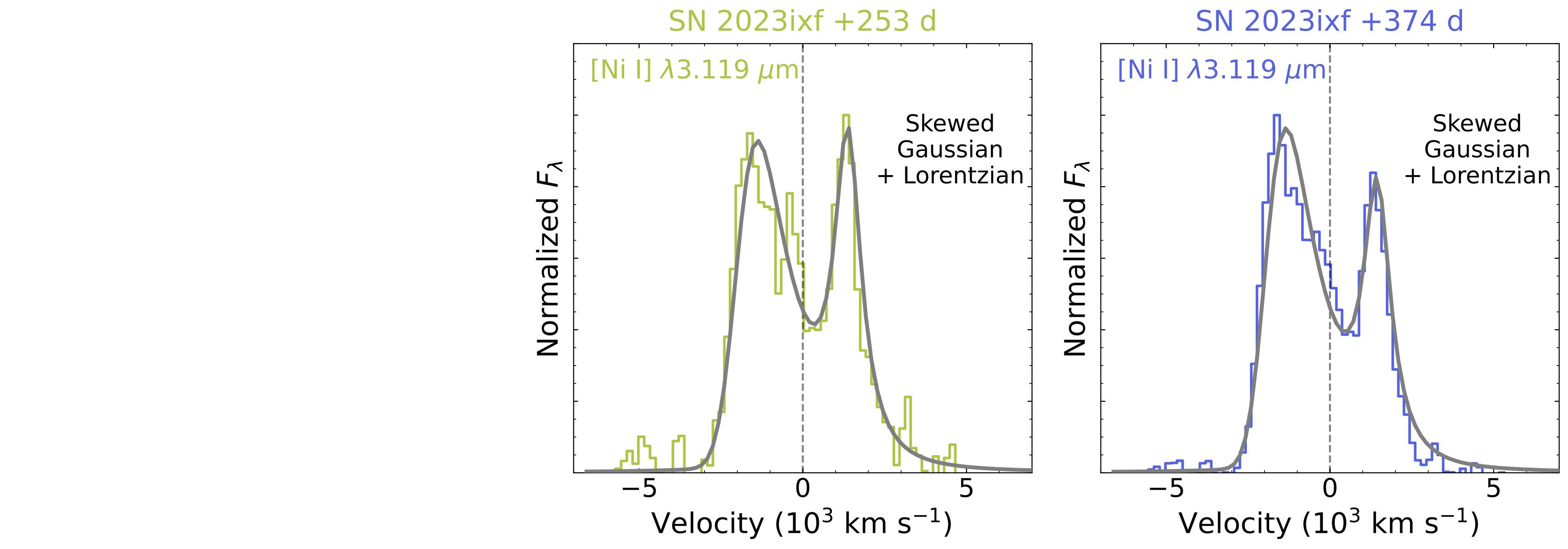}
    \caption{Line-profile fits for double-peaked \ion{Mg}{1} $1.504\ {\rm \mu m}$, \ion{Na}{1} $2.206\ {\rm \mu m}$, and [\ion{Ni}{1}] $3.12\ {\rm \mu m}$ emission lines in SN~2023ixf at different epochs shown in solid gray. 
    The line models for \ion{Mg}{1} $1.504\ {\rm \mu m}$ and the nearby \ion{Mg}{1} $1.488\ {\rm \mu m}$ are shown in dashed black. 
    We display the similarly double-peaked profile of \ion{Mg}{1} $3.867\ {\rm \mu m}$ at $+253$ d as an inset in the top-center panel.
    All three emission lines are well fit by two skewed Gaussians, which likely indicate a bipolar distribution of material or two distinct dense clumps within the metal-rich inner ejecta.}
    \label{fig:double_peaks}
\end{figure*}

The presence of a clear double-peaked line profile was first reported for \ion{Na}{1} D by \cite{Li_2025}. However, its strong P~Cygni absorption component, combined with a possible scattering component from \ion{He}{1}, complicates the interpretation of the emission component.
Similarly, the optical [\ion{Fe}{2}] $\lambda 7155$ line also shows evidence for a double-peaked profile, but blending with several nearby emission lines makes identification of the redshifted peak difficult (see Section~\ref{sec:ca_fits}).
In addition to \ion{Na}{1} D and [\ion{Fe}{2}] $\lambda 7155$, clear double-peaked profiles are observed in emission from neutral transitions of IMEs and IGEs: \ion{Mg}{1} $1.504\ {\rm \mu m}$, \ion{Na}{1} $2.206\ {\rm \mu m}$, and [\ion{Ni}{1}] $3.12\ {\rm \mu m}$.
At $+253$ d, \ion{Mg}{1} $3.867\ {\rm \mu m}$ also exhibits a double-peaked profile but fades by $+374$ d, which we show as an inset in the middle top panel of Figure~\ref{fig:double_peaks}.
We find that fitting these features with two simple Gaussian components is insufficient to reproduce the bulk of the emission, particularly owing to the blue-skewed nature of the blueshifted peaks and their steep wings. Instead, a skewed Gaussian provides a better representation of the blueshifted emission component. 

For \ion{Mg}{1} $1.504\ {\rm \mu m}$, we simultaneously fit the nearby \ion{Mg}{1} $1.488\ {\rm \mu m}$ emission by constraining pairs of line-profile components to be separated by $\Delta v \approx 3191\ {\rm km\ s^{-1}}$. 
Despite minor discrepancies between the fits and the data, two skewed Gaussian components provide a satisfactory fit to both \ion{Na}{1} $2.206\ {\rm \mu m}$ and \ion{Mg}{1} $1.504\ {\rm \mu m}$.
For [\ion{Ni}{1}] $3.12\ {\rm \mu m}$, we adopt a combination of a skewed Gaussian and a Lorentzian profile to reproduce the narrow redshifted peak. 
The resulting fits for \ion{Na}{1}, \ion{Mg}{1}, and [\ion{Ni}{1}] are shown in Figure~\ref{fig:double_peaks}.
At $+374$ d, the \ion{Na}{1} profile is notably less well constrained, likely owing to the lower spectral resolution of the JWST spectrum and the intrinsically weak emission at this epoch. 
The blueshifted peak of [\ion{Ni}{1}] $3.12\ {\rm \mu m}$ exhibits a sharper skew than can be fully captured by a skewed Gaussian without overestimating the red wing, although the overall morphology is still reasonably reproduced.

\subsection{Symmetric IME and IGE Emission?} 
\label{sec: symmetric_fits}

\begin{figure*}[t!]
    \centering
    \includegraphics[width=\textwidth]{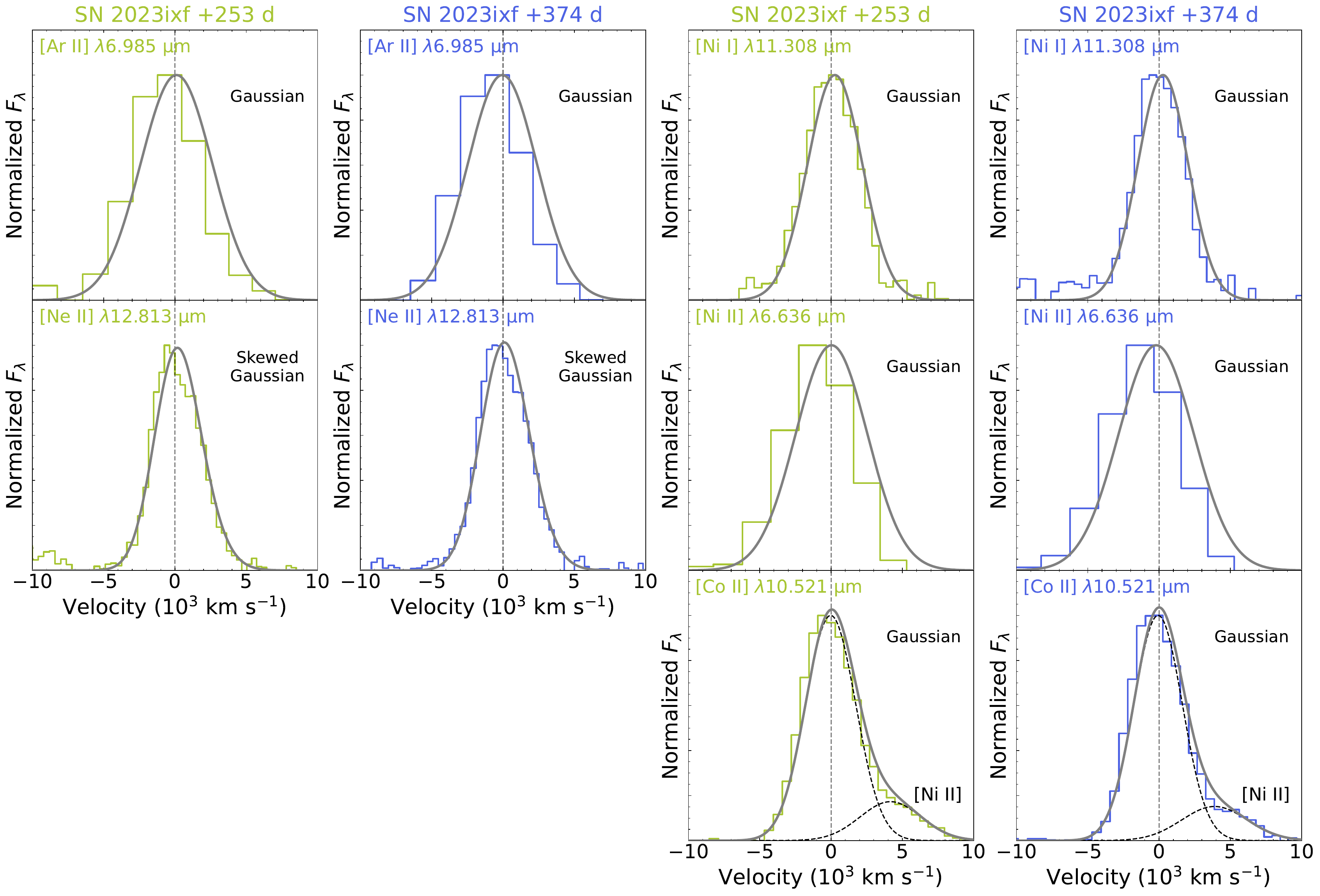}
    \caption{Line-profile fits for broad and symmetric emission from IMEs ([\ion{Ar}{2}] $6.985\ {\rm \mu m}$ and [\ion{Ne}{2}] $12.813\ {\rm \mu m}$) and IGEs ([\ion{Ni}{1}] $11.308\ {\rm \mu m}$, [\ion{Ni}{2}] $6.636\ {\rm \mu m}$, and [\ion{Co}{2}] $10.521\ {\rm \mu m}$) in SN~2023ixf at $+253$ d and $+374$ d shown in solid gray. 
    All of these emission lines are well fit by a single Gaussian profile centered near rest velocity, but the derived parameters are subject to significant uncertainty owing to the poor resolution of MIRI/LRS.
    The slightly red-skewed peak of [\ion{Ne}{2}] $12.813\ {\rm \mu m}$ is better matched by a skewed Gaussian profile.
    If emission lines from these (mostly) singly-ionized IMEs and IGEs are not artifacts of resolution effects, they may reflect density and temperature differences in the ejecta. 
    }
    \label{fig:symmetric_lines}
\end{figure*}

Several other emission lines from IMEs and IGEs exhibit profiles that are more symmetric and less offset compared to the double-peaked \ion{Na}{1} $2.206\ {\rm \mu m}$, \ion{Mg}{1} $1.504\ {\rm \mu m}$, and [\ion{Ni}{1}] $3.12\ {\rm \mu m}$ emission. 
These features mostly arise from ionized species, including [\ion{Ar}{2}] $6.985\ {\rm \mu m}$, [\ion{Ne}{2}] $12.813\ {\rm \mu m}$, [\ion{Ni}{2}] $6.636\ {\rm \mu m}$, [\ion{Ni}{1}] $11.308\ {\rm \mu m}$, and [\ion{Co}{2}] $10.521\ {\rm \mu m}$, all of which display approximately Gaussian  shapes.  
Although the detailed profiles are not well constrained owing to the low spectral resolution of MIRI/LRS, modeling these features still provides useful upper limits on the line-of-sight extent of their emitting regions. 
To account for instrumental broadening, we add the model FWHM in quadrature with the instrumental FWHM.
For [\ion{Co}{2}] $10.521\ {\rm \mu m}$, we simultaneously fit the nearby [\ion{Ni}{2}] $10.682\ {\rm \mu m}$ line by constraining its parameters relative to [\ion{Ni}{2}] $6.636\ {\rm \mu m}$, allowing an additional uncertainty of $\pm 500\ {\rm km\ s^{-1}}$ in both FWHM and velocity offset. 
The resulting fits, based on a single Gaussian profile, are shown at $+253$ d and $+374$ d in Figure~\ref{fig:symmetric_lines}.

After correcting for instrumental resolution, single Gaussian fits yield ${\rm FWHM} \approx 4600\ {\rm km\ s^{-1}}$ and $\mu \approx 100\ {\rm km\ s^{-1}}$ for [\ion{Ar}{2}] $6.985\ {\rm \mu m}$, and ${\rm FWHM} \approx 3800\ {\rm km\ s^{-1}}$ and $\mu \approx 300\ {\rm km\ s^{-1}}$ for [\ion{Ne}{2}] $12.813\ {\rm \mu m}$, but leave a slight flux excess on the blue side and a deficit on the red side of [\ion{Ne}{2}]. 
Motivated by this mild red-skewness, we alternatively fit [\ion{Ne}{2}] with a skewed Gaussian profile, which provides a modestly improved match to the observed shape.
The profiles of [\ion{Ni}{1}] $11.308\ {\rm \mu m}$ and [\ion{Ni}{2}] $6.636\ {\rm \mu m}$ exhibit similar widths, with ${\rm FWHM} \approx 4500\ {\rm km\ s^{-1}}$, while [\ion{Co}{2}] $10.521\ {\rm \mu m}$ is somewhat narrower at ${\rm FWHM} \approx 4000\ {\rm km\ s^{-1}}$. 
Taken together, these symmetric profiles seen in ionized IME and IGE lines drastically contrast with the double-peaked profiles of generally neutral species described in Section~\ref{sec:double_peaked}, most notably [\ion{Ni}{1}] $11.308\ {\rm \mu m}$.
It is suspicious that neutral emission from [\ion{Ni}{1}] exhibits two distinct profile shapes (double-peaked for [\ion{Ni}{1}] $3.12\ {\rm \mu m}$ and Gaussian for [\ion{Ni}{1}] $11.308\ {\rm \mu m}$).
If these (mostly) singly ionized IMEs and IGEs truly follow more symmetric emitting regions, they reflect potential stratification in density, temperature, or ionization within the ejecta.
However, we show in Appendix~\ref{app:miri_res} it is possible that all of these symmetric profiles are smoothed out by resolution effects, which cannot be disentangled.

\subsection{Blueshifted O Emission}
\label{sec: O_fits}

\begin{figure*}[t!]
    \centering
    \includegraphics[width=.9\textwidth]{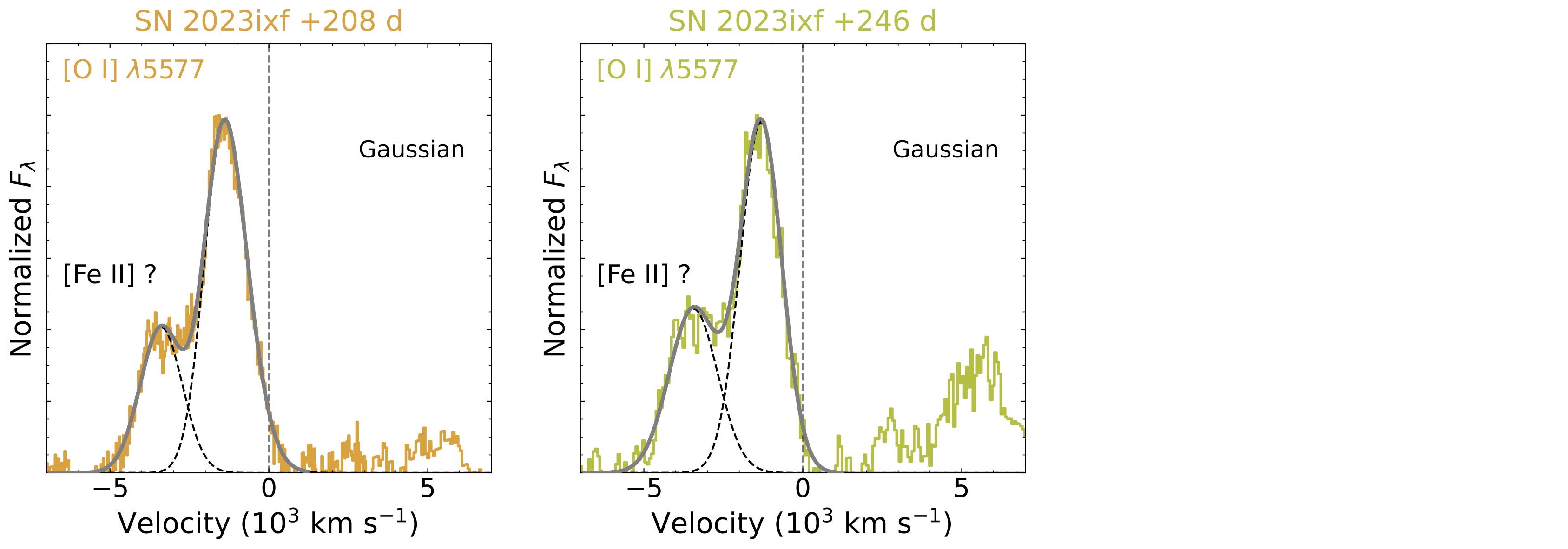}
    \includegraphics[width=.9\textwidth]{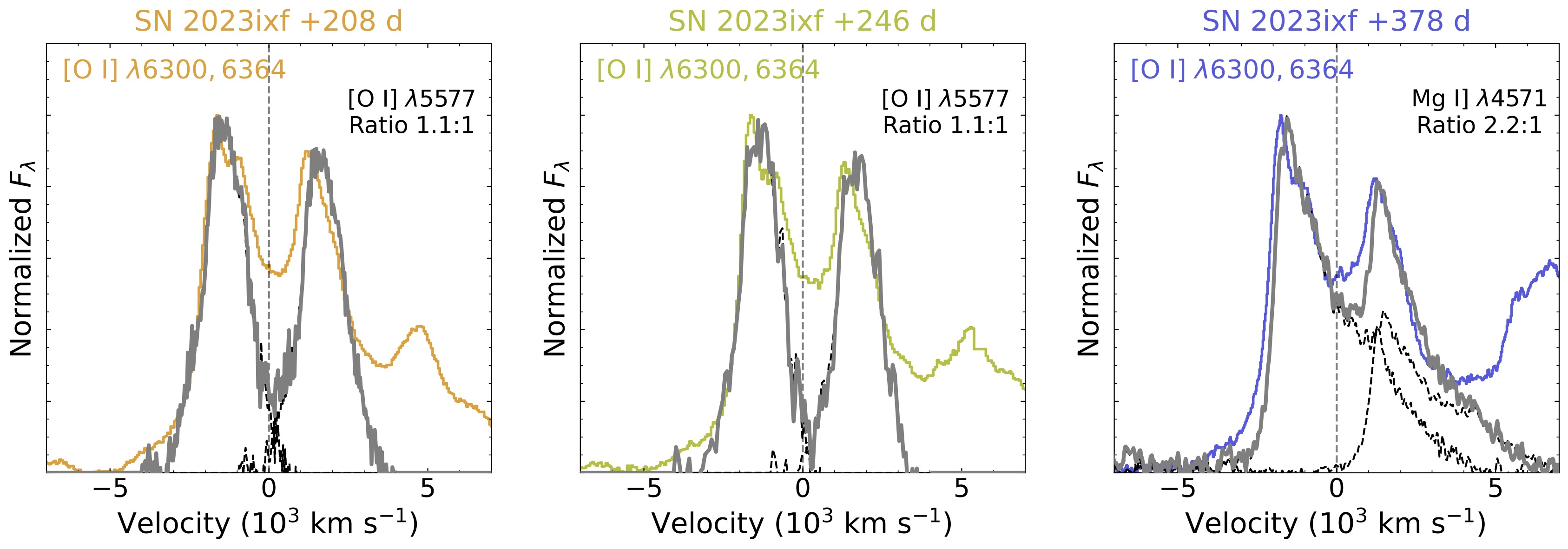}
    \includegraphics[width=.9\textwidth]{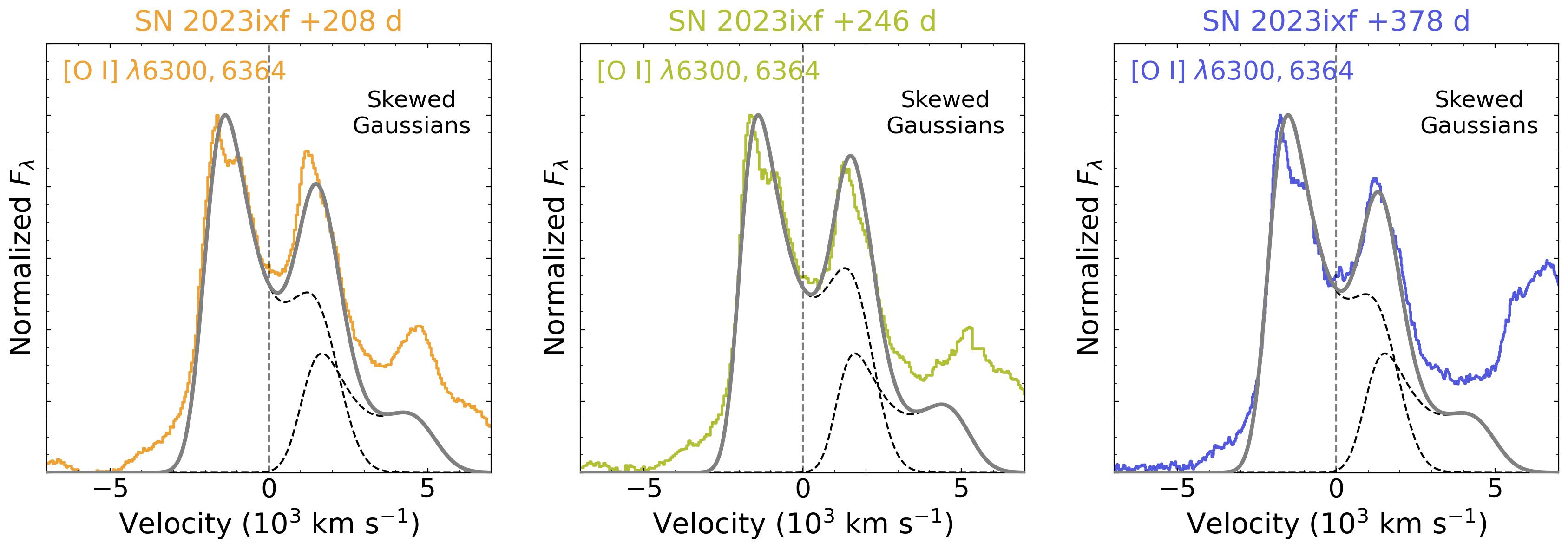}
    \caption{Line-profile fits for [\ion{O}{1}] $\lambda 5577$ and [\ion{O}{1}] $\lambda $6300, 6364 in SN~2023ixf at different epochs. 
    The models for individual lines are shown in dashed black, and the composite fits are in solid gray.
    The [\ion{O}{1}] $\lambda 5577$ emission is well matched by a single Gaussian, with an additional blueshifted Gaussian to capture possible contamination from [\ion{Fe}{2}].
    Using the [\ion{O}{1}] $\lambda 5577$ and \ion{Mg}{1}] $\lambda 4571$ profiles as templates, the symmetric peaks in [\ion{O}{1}] $\lambda $6300, 6364 can be reproduced with an optically thick (1.1:1 ratio; middle-left and center panels) and an almost optically thin (2.2:1 ratio; middle-right panel) O-rich emitting region, respectively.
    Alternatively, using the skewed Gaussians derived from \ion{Mg}{1} $1.504\ {\rm \mu m}$,  [\ion{O}{1}] $\lambda $6300, 6364 is consistent with two independent optically thin O-rich emitting regions.
    The persistently blueshifted nature of [\ion{O}{1}] $\lambda 5577$ and [\ion{O}{1}] $\lambda \lambda$6300, 6364 suggests the O-rich ejecta is intrinsically asymmetric.
    }
    \label{fig:OI_models}
\end{figure*}

The two strongest and relatively isolated O emission features are [\ion{O}{1}] $\lambda 5577$ and [\ion{O}{1}] $\lambda\lambda $6300, 6364, both of which exhibit blueshifted peaks.
For [\ion{O}{1}] $\lambda 5577$, we fit the region spanning $\sim 5400$--5600 \AA\ using two Gaussian profiles, following the procedure of \cite{Jerkstrand_2014}, to account for potential contamination from [\ion{Fe}{2}]. 
At $+208$ d, the line is well described by a single Gaussian with ${\rm FWHM} \approx 1700\ {\rm km\ s^{-1}}$ and $\mu \approx -1400\ {\rm km\ s^{-1}}$. 
The line evolves only modestly by $+246$ d, with ${\rm FWHM} \approx 1600\ {\rm km\ s^{-1}}$ and $\mu \approx -1300\ {\rm km\ s^{-1}}$. 
These fits at $+208$ d and $+246$ d are shown in the top panel of Figure~\ref{fig:OI_models}.
Based on radiative-transfer models, \citet{Jerkstrand_2015a} attributed the early blueshift in [\ion{O}{1}] $\lambda 5577$ to significant line opacity, which preferentially suppresses emission from the receding side of the ejecta. 
As the ejecta expand and become more optically thin, this blueshift is expected to diminish. 
However, in SN~2023ixf, [\ion{O}{1}] $\lambda 5577$ remains strongly blueshifted, decreasing only slightly from $\approx -1400\ {\rm km\ s^{-1}}$ at $+139$ d, the earliest epoch with a sufficiently strong detection, to $\approx -1300\ {\rm km\ s^{-1}}$ at $+349$ d, before disappearing entirely.
The persistence of this blueshift suggests that the [\ion{O}{1}] $\lambda 5577$ emission arises from a relatively dense and geometrically offset O-rich region in the ejecta.
A number of other IME lines exhibit a similar persistent blueshift.

As the SN evolves, the intensity ratio of [\ion{O}{1}] $\lambda$6300/$\lambda$6364 is expected to increase from unity in the optically thick limit to an asymptotic value of 3 in the optically thin limit under the assumption of local thermodynamic equilibrium (LTE; \citealt{Leibundgut_1991, Li_McCray_1992, Williams_1994, Jerkstrand_2017}).
Compared to [\ion{O}{1}] $\lambda 5577$, the [\ion{O}{1}] $\lambda\lambda $6300, 6364 doublet exhibits a prominent broad component centered near zero velocity, reinforcing the idea that [\ion{O}{1}] emission arises from multiple O-rich regions.
A central flux deficit becomes evident when we adopt the [\ion{O}{1}] $\lambda 5577$ profile as a template, shifted by $\Delta v \approx 3046\ {\rm km\ s^{-1}}$ (corresponding to $\Delta\lambda = 64\ {\rm \AA}$ relative to 6300 \AA), and combined in a 1.1:1 ratio to reproduce the observed amplitude of [\ion{O}{1}] $\lambda\lambda $6300, 6364, following \cite{Milisavljevic_2010}. 
The result from this scaling is consistent with emission from a single optically thick O-rich region at $+208$ d and $+246$ d (see Figure~\ref{fig:OI_models}).

Alternatively, stellar-evolution models predict that Mg and O occupy similar spatial regions in the pre-explosion core (\citealt{Mazzali_2005,Maeda_2006}).
In this case, the line profiles of isolated Mg and O transitions should closely resemble one another, as commonly observed in SESNe (\citealt{Modjaz_2008, Taubenberger_2009, Milisavljevic_2010}). 
This expectation is supported here by the shared blueshifted emission peaks of [\ion{O}{1}] and multiple \ion{Mg}{1} lines (see Figure~\ref{fig:line_profiles}). 
\cite{Kumar_2025} have already explored the use of \ion{Mg}{1}] $\lambda 4571$ as a template for modeling [\ion{O}{1}] $\lambda\lambda $6300, 6364. 
We perform a similar empirical test by superimposing two continuum-subtracted \ion{Mg}{1}] $\lambda 4571$ profiles separated by $\Delta v \approx 3046\ {\rm km\ s^{-1}}$, summed, and rescaled to match the observed amplitude of the [\ion{O}{1}] profile at $+378$ d. 
This epoch is chosen because \ion{Mg}{1}] $\lambda 4571$ is expected to be relatively isolated and optically thin ($\gtrsim +250$ d; \citealt{Jerkstrand_2015a}), while [\ion{O}{1}] $\lambda 5577$ has already faded by around $+349$ d. 
An example fit is shown in the middle-right panel of Figure~\ref{fig:OI_models}.
We find that a 2.2:1 combination of the \ion{Mg}{1}] $\lambda 4571$ template provides a reasonable match to the observed [\ion{O}{1}] $\lambda\lambda $6300, 6364 profile, suggesting a more optically thin regime. 
However, a residual deficit near zero velocity remains, which may reflect a higher optical depth in \ion{Mg}{1}], additional emission from a more extended and volume-filling O-rich region, or the effects of dust extinction.

The fact that \ion{Mg}{1} and \ion{Na}{1} exhibit double-peaked profiles while [\ion{O}{1}] does not is noteworthy. 
One possibility is that a redshifted emission peak is present but significantly attenuated by ejecta dust and fades below the CSM horns of H$\alpha$.
Using the best-fit skewed Gaussian components for \ion{Mg}{1} $1.504\ {\rm \mu m}$ as line templates (see Section~\ref{sec:double_peaked}), we perform the same addition and scaling, assuming both components are optically thin. 
These fits are shown in the bottom row of Figure~\ref{fig:OI_models}.
At $+208$ d, the model underestimates the peak amplitude of the [\ion{O}{1}] $\lambda 6364$ component, indicating that the emission is not yet fully optically thin. 
By contrast, the $+246$ d and $+378$ d [\ion{O}{1}] $\lambda\lambda $6300, 6364 profiles are well-reproduced by two optically thin O-emitting regions.
The extended blue wing ($-5000\ {\rm km\ s^{-1}}\lesssim v\lesssim -3000\ {\rm km\ s^{-1}}$) and possibly any red wing hindered by H$\alpha$ could come from O mixed in the H envelope.

\subsection{Multicomponent [\ion{Ca}{2}] Emission}
\label{sec:ca_fits}

In SNe II, \ion{Ca}{2} emission may arise from two spatially distinct regions within the ejecta: primordial Ca in the H-rich envelope (\citealt{Li_McCray_1993, Kozma_Fransson_1998, Jerkstrand_2012, Michel_2025}) or a more centrally concentrated component produced by explosive nucleosynthesis (\citealt{Jerkstrand_2015a, Dessart_2021, Fang_2024}).
Because the two lines of the [\ion{Ca}{2}] doublet share a common lower level (the ground state of \ion{Ca}{2}) and have similar transition probabilities (\citealt{Osterbrock_1951, Warner_1968, Black_1972, Li_McCray_1993}), their intensity ratio in the optically thin limit in LTE is set by the ratio of the statistical weights of their upper levels, $I_{7291}/I_{7324}=3/2$ (\citealt{Spyromilio_1993}).
Motivated by the presence of both blue and red flux excesses near zero velocity, which we associate with [\ion{Ca}{2}] (see Section~\ref{sec:oxygen_magnesium}), we explore two geometric configurations: (1) a central narrow Gaussian component combined with a broader asymmetric shell, and (2) three distinct Gaussian components. 
The latter configuration is partly motivated by the double-peaked profiles observed in \ion{Na}{1}, \ion{Mg}{1}, and [\ion{Ni}{1}] (see Section~\ref{sec:double_peaked}).
For both configurations, we fix the separation between the doublet components to $\Delta v \approx 1358\ {\rm km\ s^{-1}}$, corresponding to a wavelength separation of $\Delta\lambda = 33\ {\rm \AA}$.

Here, we apply the modeling framework presented by \cite{Jerkstrand_2015b} to characterize the emission complex in the wavelength range 7000--7600\ {\rm \AA}, which has been applied to several CCSNe, including SN~2023ixf, in previous studies (e.g., \citealt{Terreran_2016, Gutierrez_2020, Sollerman_2021,Prentice_2022, Michel_2025}).
The spectral features considered in this regime are [\ion{Ca}{2}] $\lambda\lambda 7291$, 7324, [\ion{Fe}{2}] $\lambda\lambda 7155$, 7172, [\ion{Ni}{2}] $\lambda\lambda 7378$, 7412, and [\ion{Fe}{2}] $\lambda\lambda 7388$, 7453. 
The relative intensity ratios for [\ion{Fe}{2}] and [\ion{Ni}{2}] are adopted from \cite{Jerkstrand_2015b}. 
We use the resolution-corrected FWHM and velocity offset measured from a Gaussian fit to the MIR [\ion{Ni}{2}] $6.636\ {\rm \mu m}$ line to constrain the [\ion{Ni}{2}] $\lambda\lambda 7378$, 7412 doublet.
As no relatively isolated [\ion{Fe}{2}] emission is identified, we assume for simplicity that all [\ion{Fe}{2}] lines in this region share a Gaussian profile with the same FWHM and velocity offset as the MIR [\ion{Co}{2}] $10.521\ {\rm \mu m}$ emission. 
This choice is motivated by the expectation that [\ion{Fe}{2}] and [\ion{Co}{2}] trace similar spatial distributions within the ejecta.
We allow an additional uncertainty of $\pm 500\ {\rm km\ s^{-1}}$ in both FWHM and velocity offset for all lines constrained using NIR and MIR templates.

\begin{figure*}[t!]
    \centering
    \includegraphics[width=.9\textwidth]{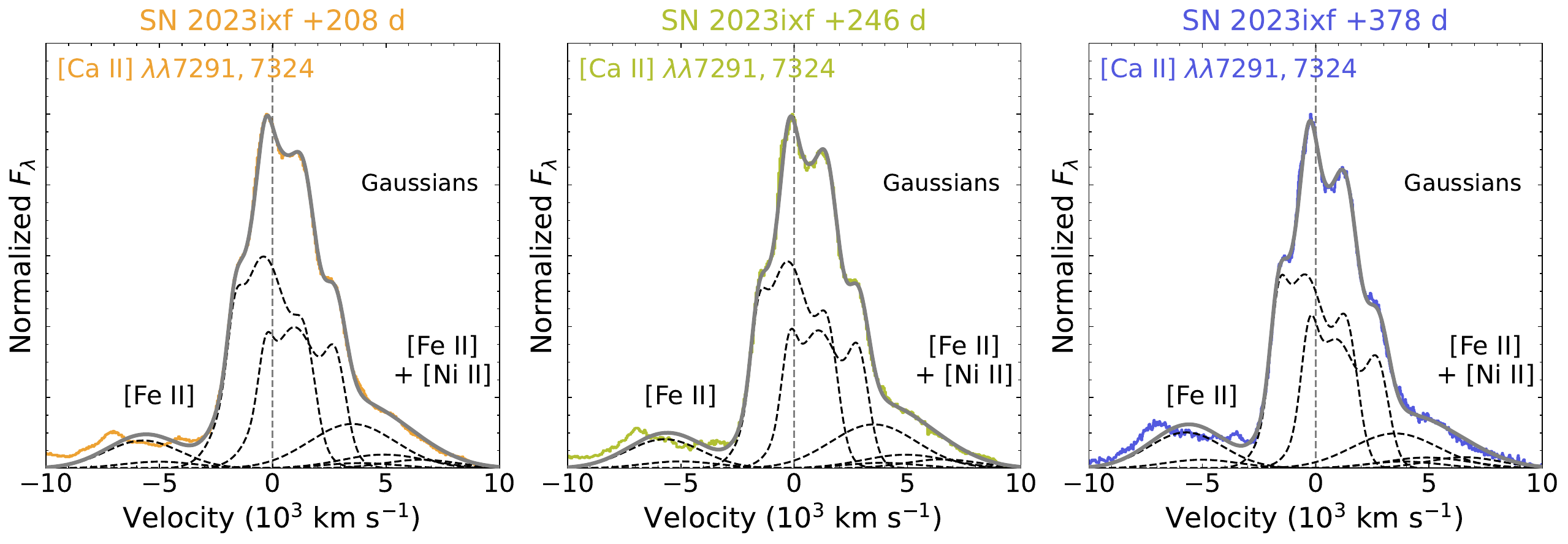}
    \caption{Line-profile fits for [\ion{Ca}{2}] $\lambda\lambda7291$, 7324 in SN~2023ixf at three different epochs. 
    The fits for individual lines are shown in dashed black, and the composite fits are in solid gray. 
    The [\ion{Ca}{2}] emission is heavily blended with several weak lines from [\ion{Fe}{2}] and [\ion{Ni}{2}]. 
    Each line in the [\ion{Ca}{2}] $\lambda\lambda 7291$, 7324 doublet can be fit with three distinct Gaussians, consistent with contributions from both the innermost ejecta and possibly Ca present close to the outer envelope.}
    \label{fig:CaII_models}
\end{figure*}

In either geometric configuration, the central Gaussian component remains relatively unchanged and optically thin throughout, with an intensity ratio of $I_{7291}/I_{7324} = 3/2$, consistent with theoretical expectations.
In the asymmetric-shell configuration, the broad component shows only a modest evolution in width from ${\rm FWHM} \approx 2700\ {\rm km\ s^{-1}}$ at $+208$ d to ${\rm FWHM} \approx 2500\ {\rm km\ s^{-1}}$ at $+378$ d.
The central cavity associated with the asymmetric shell also remains stable over time, with $v_{\rm in} \approx 1600\ {\rm km\ s^{-1}}$ and $v_c \approx -100\ {\rm km\ s^{-1}}$ across all three epochs. 
In contrast with the central Gaussian core, the broad asymmetric-shell components evolve from being marginally optically thick ($I_{7291}/I_{7324}\approx 0.93$ at $+208$ d) to becoming increasingly optically thin ($I_{7291}/I_{7324}\approx 0.77$ by $+378$ d).
On the other hand, the redshifted components grow progressively optically thin in the multi-Gaussian configuration, while the blueshifted component remains optically thin throughout, possibly indicating a denser Ca-emitting region away from our line of sight.

We also attempted to fit [\ion{Ca}{2}] using two skewed Gaussian components, following the approach adopted for other double-peaked emission lines. 
However, we find that no physically reasonable combination of parameters is able to reproduce the observed line profiles.
Given the multipeaked nature in other IMEs and also several H transitions, we favor and show only the best-fit [\ion{Ca}{2}] profiles with multiple Gaussians in Figure~\ref{fig:CaII_models}.
Notably, while the red wing of the emission complex around [\ion{Ca}{2}] is well matched by four Gaussian-shaped [\ion{Fe}{2}] and [\ion{Ni}{2}] emission lines, the blue wing has two peaks around the zero velocity of [\ion{Fe}{2}] $\lambda 7155$.
This could mean that emission from [\ion{Fe}{2}] also possesses double-peaked profiles.
On the other hand, telluric absorption by O$_2$ (B band) near this wavelength range introduce significant uncertainties in the continuum placement, as well as the inferred amplitudes and profiles of the [\ion{Fe}{2}] $\lambda\lambda 7155$, 7172 lines. 
As a result, the empirical fits in this region should be regarded as approximate rather than fully physically constrained.
Although the precise line-profile structure responsible for the extended emission on either side of [\ion{Ca}{2}] remains uncertain, multiple Ca-emitting regions are clearly required to reproduce the observed profiles.

\section{Discussion}
\label{sec:discussion}

\subsection{Inferred Ejecta Structure}
\label{sec:structure}

The diversity in line profiles analyzed in Section~\ref{sec:line_profile_analysis} suggests a highly nonspherical distribution of material within the ejecta. 
Our findings are similar to the recent study of SN~2024ggi by \cite{Jacobson-Galan_2026}, another very nearby SN II with JWST coverage in the nebular phase, where similar double-peaked profiles in [\ion{Ni}{1}] and \ion{Mg}{1} were identified.
State-of-the-art 3D hydrodynamic simulations consistently predict asymmetric ejection of Ni-rich material in large-scale structures composed of several plumes. 
These plumes can evolve into Rayleigh–Taylor instability (RTI) ``fingers" as they interact with reverse shocks developed at compositional interfaces (\citealt{Vartanyan_2025b, Vartanyan_2025}). 
A key outcome of these simulations is that Ni-rich plumes can penetrate multiple layers with distinct compositions, reaching the base of, or even extending beyond, the H/He interface, thereby producing asymmetric heating of the ejecta (see, e.g., \citealt{Wongwathanarat_2015, Sandoval_2021}).

Within this framework, many of the line profiles analyzed in Section~\ref{sec:line_profile_analysis} can be understood as the combined result of an asymmetric explosion geometry and stratified elemental distribution. 
In particular, the double-peaked profile of [\ion{Ni}{1}] $3.12\ {\rm \mu m}$ likely arises from dense, localized (maybe even self-shielded) Ni clumps where recombination occurs more effectively (\citealt{Wang_2005, Dessart_2018}), while emission from singly ionized IGEs such as [\ion{Ni}{2}] and [\ion{Co}{2}] trace warmer, more diffuse, and volume-filling regions associated with cavities carved out by the nickel-bubble effect. 
Emission from singly ionized IMEs ([\ion{Ar}{2}] and [\ion{Ne}{2}]) then trace a region close to the nickel bubbles (or potentially mixed within) to achieve their higher ionization potentials (\citealt{Kotak_2006,Dessart_2025b}), consistent with their comparable FWHMs to singly ionized IGEs.
By contrast, emission from neutral IMEs such as \ion{Mg}{1} and \ion{Na}{1}, which primarily form via recombination (\citealt{Jerkstrand_2015a}), likely originates in compressed regions around these Ni-rich cavities. 

The strong evidence for intrinsic ejecta asphericities inferred from our line-profile modeling is consistent with late-time spectropolarimetric observations of SN~2023ixf (up to $\sim 120$ days post-explosion; \citealt{Singh_2024, Shrestha_2025, Vasylyev_2026}), which showed an increase in polarization as the SN transitions from the photospheric phase to the nebular phase.
\cite{Vasylyev_2026} argued that the spectropolarimetric evolution of SN~2023ixf can be explained by a bipolar explosion, a scenario commonly invoked for strongly interacting SNe exhibiting high polarization signals (\citealt{Wang_2001, Wang_2002, Dessart_2021b, Bilinski_2024, Vasylyev_2024}). 
Several independent studies of SN~2023ixf have reached similar conclusions based on comparisons of the [\ion{O}{1}] $\lambda\lambda $6300, 6364 and [\ion{Ca}{2}] $\lambda\lambda 7291$, 7324 line profiles, suggesting an axisymmetric structure in which an O-rich torus surrounds bipolar lobes of explosively synthesized material ejected along the jet direction (\citealt{Ferrari_2024, Fang_2025b}). 
A comparable interpretation was proposed by \cite{Jacobson-Galan_2026} for SN~2024ggi, where pronounced polar enhancements are required to reproduce the observed IGE and IME line profiles.
While an axisymmetric $^{56}$Ni distribution in SN~2023ixf is plausible, 3D simulations indicate that the number, morphology, and orientation of dominant Ni plumes develop turbulently, rather than conforming to a bipolar distribution with axial symmetry (\citealt{Utrobin_2017, Burrows_2020, Wang_Burrows_2024, Vartanyan_2025b, Vartanyan_2025}). 

Alternatively, some studies suggest that late-time asymmetries in the Ni-rich ejecta are seeded by perturbations in the neutrino-heated bubble layer and RTI at shell interfaces (e.g., \citealt{Wongwathanarat_2015, Giudici_2025}); however, this mechanism also does not inherently produce a bipolar configuration.
Therefore, we adopt a more conservative interpretation in which the Ni-rich ejecta are dominated by at least two large plumes that are not necessarily aligned along any axis, perhaps with one directed away from the line of sight. 
We note that mechanisms positing a jet-driven origin for SN~2023ixf (\citealt{Reynoso_2024,Soker_2025}) can also explain a bipolar ejecta, as predicted by 3D magneto-hydrodynamic simulations (e.g., \citealt{Mosta_2018,Obergaulinger_2021}).
Ultimately, fully constraining the ejecta morphology in SN~2023ixf will require future 3D simulations that follow CCSN evolution into the nebular phase and incorporate detailed spectral synthesis to match observations.

\subsection{The CSM Geometry}
\label{sec:csm}

As discussed in detail in Section~\ref{sec:Halpha} and previous studies (e.g., \citealt{Singh_2024,Jacobson-Galan_2025,Kumar_2025,Bostroem_2025, Dessart_2026}), the broad CSM horns shared by H and He emission features likely result from reprocessing of X-rays by the CDS and photoionization of the fast-moving ejecta. 
Through our profile decomposition in Section~\ref{sec:H_profiles}, however, we have reasons to suspect that an additional intermediate-width component ($-4000\ {\rm km\ s^{-1}}\lesssim v\lesssim 4000\ {\rm km\ s^{-1}}$) exists and is related to the dense --- and likely aspherical --- CSM that caused the early-time narrow features (\citealt{Jacobson-Galan_2023,Bostroem_2023,Smith_2023}) and high polarization signals (\citealt{Vasylyev_2023,Vasylyev_2026,Singh_2024,Shrestha_2025}).
Any disk-like or torus-like dense CSM around SN~2023ixf's progenitor must have had a limited radial extent given the lack of P~Cygni absorption from the unshocked CSM and the early disappearance of narrow features (\citealt{Smith_2023,Dickinson_2025}).
In this scenario, the fast ejecta engulf the CSM, and any sustained interaction signatures are hidden below the optically thick photosphere. 
As the photosphere recedes, the post-shock CSM emerges and contributes intermediate-width emission from the ejecta crashing into the post-shock gas (see \citealt{Smith_2015}).

\begin{figure*}[t!]
    \centering
    \includegraphics[width=.8\textwidth]{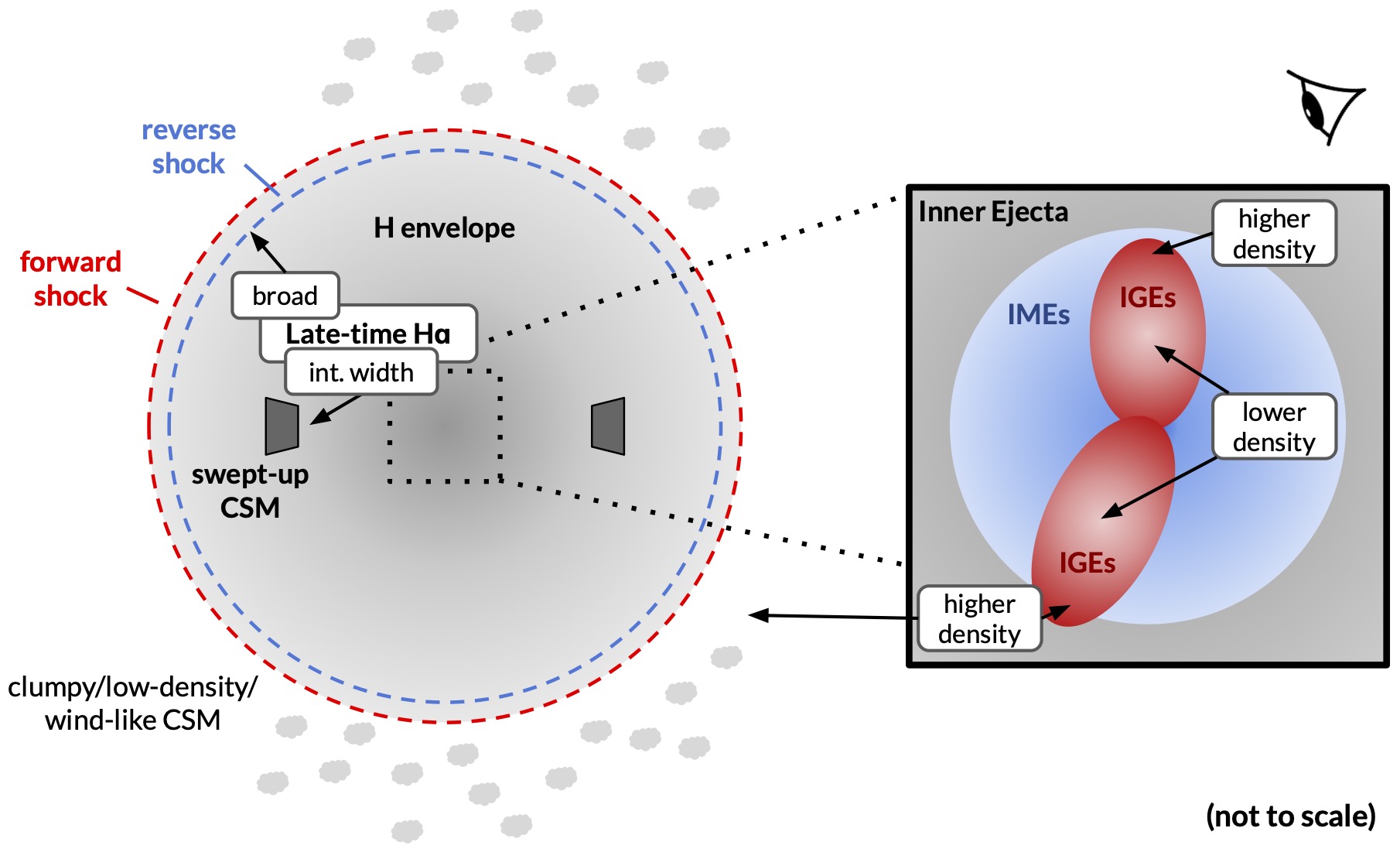}
    \caption{The inferred ejecta and CSM structure for SN~2023ixf from our line-profile fits.
    Various shock-powered emission lines indicate the presence of an intermediate-width component from the ejecta crashing into swept-up, aspherical CSM. 
    Various double-peaked profiles indicate that IMEs and IGEs follow a bipolar distribution, but do not have to follow axial symmetry along the poles.}
    \label{fig:ejecta_structure}
\end{figure*}

Such an emergence of underlying emission from the ejecta interacting with an aspherical CSM has also been demonstrated by 2D radiation-hydrodynamic simulations (\citealt{Kurfurst_2020,Kurfurst_2026}).
By $+749$ d, it is likely that all of the slowest moving CSM is already swept up by the shock, given the lack of narrow emission. 
The swept-up CSM disk is another location in addition to the ejecta that is cool and dense enough for dust growth, which would explain the lack of a redshifted peak in the intermediate-width component.
Together with our interpretations for an asymmetric ejecta, we show our inferred ejecta and CSM composite structure for SN~2023ixf in Figure~\ref{fig:ejecta_structure}.
The picture described here is essentially a combination of the respective evolution suggested for PTF11iqb and iPTF14hls (\citealt{Smith_2015,Andrews_Smith_2018}).

\subsection{Effects of Dust}
\label{sec:dust}

The presence of dust within the ejecta of CCSNe has a profound impact on the interpretation of ejecta geometry based on nebular line-profile modeling alone.  
Visual-wavelength emission-line signatures of dust formation in CCSNe were first firmly established in the case of SN~1987A \citep{danziger89,Lucy_1989}, identified as a progressive blueshift in the [\ion{O}{1}] $\lambda\lambda$6300, 6364 and [\ion{C}{1}] $\lambda\lambda9824$, 9850 profiles.  
These were attributed to attenuation of emission from the receding side of the ejecta. 
Similar behavior has since been reported in numerous SNe (e.g., \citealt{Kotak_2009, Maguire_2010, smith08jc, Smith_2012,Rho_2018,eliasrosa18,sa20,dickinson24,Zsiros_2024}), establishing red-wing suppression as an indicator of internal dust extinction.
In SN~2023ixf, strong line-profile asymmetries are clearly present, particularly in multiple H transitions (see Section~\ref{sec:hydrogen_helium}; \citealt{Singh_2024, Jacobson-Galan_2025, Medler_2025, Singh_2026}), as well as in optical [\ion{O}{1}] and \ion{Mg}{1}] emission. 

In addition, emission of rotational-vibrational transitions from CO was observed at $+199$ d for first-overtone bands and at $+253$ d for fundamental bands (\citealt{Medler_2025,Park_2025}).
Molecular emission signals efficient cooling of the ejecta and is often a hallmark signature prior to the onset of dust formation. 
Analyses of the thermal IR emission and radiative-transfer modeling of H$\alpha$ by \cite{Singh_2026} provided direct evidence for multicomponent dust growth (with a cold silicate component and a cold graphite component of $\sim 10^{-3}\ M_{\odot}$, each by $+723$ d). 
In particular, \cite{Singh_2026} found that a clumpy dust distribution with modest total effective dust mass can reasonably reproduce the strong blue–red asymmetry observed in H$\alpha$, in agreement with earlier studies (e.g., \citealt{Bevan_Barlow_2016}).
A similar conclusion for growing dust mass was presented by \cite{Dessart_2026} through detailed radiative-transfer simulations.

Given the possibility of newly formed dust in SN~2023ixf (\citealt{Jacobson-Galan_2025,Dessart_2026,Singh_2026}), a crucial question is whether the asymmetries inferred in Section~\ref{sec:structure} reflect intrinsic ejecta geometry, or if they arise from dust attenuation.
Attenuation of emission lines by dust should follow a strong wavelength dependence.
However, one apparent difficulty for invoking dust attenuation to explain the blueshifted line profiles in SN~2023ixf is that the observed wavelength dependence is weak.
This is especially clear when comparing the line profiles of H$\alpha$, Pa$\alpha$, and Br$\alpha$ in Figure~\ref{fig:line_profiles}, which exhibit remarkably similar levels of red-wing suppression at the epochs analyzed in Section~\ref{sec:line_profile_analysis} (but cf.~\citealt{Jacobson-Galan_2025}, who suggest a potentially stronger wavelength dependence). 
Radiative-transfer models by \cite{Dessart_2025} likewise found that ejecta dust does not introduce strong blue-red line asymmetries in optical emission lines. 
However, this absence of strong wavelength dependence can be resolved with larger grain size and different dust composition (\citealt{Bevan_Barlow_2016, Medler_2025}).
Other than H emission, there is some evidence for wavelength-dependent effects in other species, such as between \ion{Mg}{1}] $\lambda4571$ and \ion{Mg}{1} $1.504\ \mu$m and between different [\ion{O}{1}] transitions. 
Although, line-formation mechanisms may play a bigger role than dust for these metal lines.
In any case, ejecta dust in SN~2023ixf acts to suppress the redshifted peak of any intrinsically double-peaked emission originating from the innermost ejecta, which is most clearly seen in the temporal evolution of \ion{Mg}{1} $1.504\ {\rm \mu m}$.
We therefore conclude that the inner ejecta of SN~2023ixf are likely intrinsically asymmetric, while acknowledging that more self-consistent modeling is needed to disentangle the effects of dust.

\section{Conclusions}
\label{sec:conclusions}

In this work, we present extensive optical and NIR observations of SN~2023ixf in the nebular phase ($+89$ d to $+749$ d after explosion).
Together with supplemental JWST spectra, we empirically fit a few interesting emission-line profiles to constrain the structure of the ejecta and the CSM. 
We summarize our key findings below.
\begin{itemize}
\item Our optical and NIR spectral series of SN~2023ixf reveal a plethora of emission lines from atomic and ionic transitions, including typical nebular emission lines seen in other SNe II like \ion{Mg}{1}] $\lambda 4571$, [\ion{O}{1}] $\lambda\lambda $6300, 6364, H$\alpha$, [\ion{Ca}{2}] $\lambda\lambda7291$, 7324, and the \ion{Ca}{2} NIR triplet.
\item We find evidence for weak [\ion{C}{1}] $\lambda 8727$ and [\ion{C}{1}] $\lambda\lambda 9824$, 9850 emission, both of which exhibit a pronounced blueshift of $\sim -1600\ {\rm km\ s^{-1}}$ akin to the [\ion{O}{1}] $\lambda\lambda $6300, 6364 doublet. The similarity in the profiles of [\ion{C}{1}] and [\ion{O}{1}] may reflect a common origin in the ejecta and a spatially distinct emitting region for CO.
\item Using the H$\alpha$ line profile, we calculate the expected evolution of reverse-shock luminosity $L_{\rm sh}$, assuming a constant wind mass-loss rate of $\dot{M}=10^{-4}\ M_{\odot}\ {\rm yr^{-1}}$ and a wind velocity of $v_{\rm wind}=25\ {\rm km\ s^{-1}}$. 
The resulting evolution of $L_{\rm sh}$ is consistent with a radiative reverse shock and dominates the heating source over $^{56}$Ni decay by $+600$ d. 
\item The H$\alpha$ profile at an intermediate epoch, where both decay-powered and shock-powered components exist, can be reproduced by scaling and summing the H$\alpha$ profile from an earlier epoch that is mostly decay-powered and at a later epoch that is mostly shock-powered. We hypothesize that the late-time intermediate-width emission peak at $-3000\ {\rm km\ s^{-1}}$ arises from the ejecta crashing into the swept-up dense CSM that was initially aspherical (e.g., a disk or a torus).
\item We find double-peaked emission from \ion{Mg}{1} $1.504\ {\rm \mu m}$, \ion{Na}{1} $2.206\ {\rm \mu m}$, and [\ion{Ni}{1}] $3.12\ {\rm \mu m}$ across multiple epochs, suggesting an intrinsically asymmetric ejecta geometry. Emission lines from singly ionized IMEs and IGEs are generally broader and more symmetric, but this could be a resolution effect. The double-peaked profiles highlight an asymmetric ejection of Ni-rich material, with at least two large plumes, consistent with predictions from 3D simulations.
\end{itemize}

It is interesting that SN~2023ixf and SN~2024ggi, two of three SNe II with published JWST nebular-phase spectra (SN~2022acko being the third; \citealt{Medler_2026}), both exhibit double-peaked emission indicative of asymmetric ejecta.  
More MIR nebular-phase spectroscopy of SNe II with JWST and synthetic spectra calculated from 3D neutrino-driven simulations that account for all the underlying processes (e.g., molecules, dust) and differing configurations of CSM are needed to enable a more robust comparison and assess whether all SNe II are intrinsically asymmetric.

\begin{acknowledgments}
Time domain research by the University of Arizona team and D.J.S.~is supported by National Science Foundation (NSF) grants 2308181, 2407566, and 2432036. 
K.A.B.~is supported by an LSST-DA Catalyst Fellowship; this publication was thus made possible through the support of grant 62192 from the John Templeton Foundation to LSST-DA. 
L.A.K.~is supported by NASA through a Hubble Fellowship grant HF2-51579.001-A awarded by STScI, which is operated by the Association of Universities for Research in Astronomy, Inc., for NASA, under contract NAS5-26555. 
J.E.A.~is supported by the international Gemini Observatory, a program of NSF's NOIRLab, which is managed by the Association of Universities for Research in Astronomy (AURA) under a cooperative agreement with the U.S. NSF, on behalf of the Gemini partnership of Argentina, Brazil, Canada, Chile, the Republic of Korea, and the United States of America.
M.S.~acknowledges funding from the Australian Research Council (ARC) Centre of Excellence CE230100016.
The research of A.V.F. has been financially supported by many private generous donations. 
N.F.~acknowledges support from the National Science Foundation Graduate Research Fellowship Program under Grant No. DGE-2137419.
J.A.G.~acknowledges financial support from NASA grant 23-ATP23-0070.
C.P.G.~acknowledges financial support from grant RYC2024-050959-I, funded by MICIU/AEI/10.13039/501100011033 and the FSE+, as well as from projects PID2023-151307NB-I00, PIE 20215AT016, and CEX2020-001058-M, and the MaX-CSIC Excellence Award MaX4-SOMMA-ICE.
Supernova research at Rutgers University is supported in part by the NSF award AST-2407567.
S.V.~and the UC Davis time-domain research team ac- knowledge support by NSF grants AST-2407565.
X.-F.W.~is supported by the National Science Foundation of China (grants 12288102 and 12033003), the Mahua Teng Foundation, and the Tencent Xplorer Prize.
Y.Y.'s research is partially supported by the Tsinghua University Dushi Program.

The authors respectfully acknowledge that the University of Arizona is on the land and territories of Indigenous peoples. Today, Arizona is home to 22 federally recognized tribes, with Tucson being home to the O'odham and the Yaqui. The University strives to build sustainable relationships with sovereign Native Nations and Indigenous communities through education offerings, partnerships, and community service.

This work makes use of data from the Las Cumbres Observatory global telescope net- work, which is supported by NSF grant AST- -2308113. 
This paper makes use of data from the AAVSO Photometric All Sky Survey, whose funding has been provided by the Robert Martin Ayers Sciences Fund and from the NSF (AST-1412587). 
Some observations reported here were obtained at the MMT Observatory, a joint facility of the University of Arizona and the Smithsonian Institution.
This work is based in part on archival data obtained with the NASA Infrared Telescope Facility, which is operated by the University of Hawaii under a contract with the National Aeronautics and Space Administration.

The National Science Foundation's (NSF's) Kitt Peak National Observatory (KPNO) sits atop l'oligam Du'ag (Manzanita Shrub Mountain). Astronomers are honored to be permitted to conduct scientific research on the sacred mountain located in the homelands of the Schuk Toak District within the Tohono O'odham Nation. We honor their past, present, and future generations, who have lived here for time immemorial and will forever call this place home.
KPNO is a Program of NSF’s NOIRLab, which is managed by the Association of Universities for Research in Astronomy (AURA) under a cooperative agreement with the NSF.

This work was enabled by observations made from the Gemini North telescope, located within the Maunakea Science Reserve and adjacent to the summit of Maunakea. We are grateful for the privilege of observing the Universe from a place that is unique in both its astronomical quality and its cultural significance.
Observations obtained at the Gemini Observatory, which is operated by the Association of Universities for Research in Astronomy, Inc., under a cooperative agreement with the NSF on behalf of the Gemini partnership: the National Science Foundation (United States), the National Research Council (Canada), CONICYT (Chile), Ministerio de Ciencia, Tecnologa e Innovacin Productiva (Argentina), and Ministrio da Cincia, Tecnologia e Inovao (Brazil).

The LBT is an international collaboration among institutions in the United States and Europe. At the time data were acquired for this research, LBT Corporation Members were the University of Arizona on behalf of the Arizona Board of Regents; Istituto Nazionale di Astrofisica, Italy; LBT Beteiligungsgesellschaft, Germany, representing the Max-Planck Society, the Leibniz Institute for Astrophysics Potsdam, and Heidelberg University; and The Ohio State University, representing The Ohio State University, University of Notre Dame, University of Minnesota, and University of Virginia.  This research used the facilities of the Italian Center for Astronomical Archives (IA2) operated by INAF at the Astronomical Observatory of Trieste. Observations have benefited from the use of ALTA Center (alta.arcetri.inaf.it) forecasts performed with the Astro-Meso-Nh model. Initialization data of the ALTA automatic forecast system come from the General Circulation Model (HRES) of the European Centre for Medium Range Weather Forecasts.

A major upgrade of the Kast spectrograph on the Shane 3\,m telescope at Lick Observatory, led by Brad Holden, was made possible through gifts from the Heising-Simons Foundation, William and Marina Kast, and the University of California Observatories. 
Research at Lick Observatory is partially supported by a gift from Google.

Some of the data presented herein were obtained at Keck Observatory, which is a private 501(c)3 nonprofit organization operated as a scientific partnership among the California Institute of Technology, the University of California, and the National Aeronautics and Space Administration (NASA). The Observatory was made possible by the generous financial support of the W.~M.~Keck Foundation. The authors wish to recognize and acknowledge the very significant cultural role and reverence that the summit of Maunakea has always had within the Native Hawaiian community. We are most fortunate to have the opportunity to conduct observations from this mountain.

Funding for SDSS-III has been provided by the Alfred P. Sloan Foundation, the Participating Institutions, the NSF, and the U.S. Department of Energy Office of Science. The SDSS-III website is \url{http://www.sdss3.org/}. SDSS-III is managed by the Astrophysical Research Consortium for the Participating Institutions of the SDSS-III Collaboration, including the University of Arizona, the Brazilian Participation Group, Brookhaven National Laboratory, Carnegie Mellon University, University of Florida, the French Participation Group, the German Participation Group, Harvard University, the Instituto de Astrofisica de Canarias, the Michigan State/Notre Dame/JINA Participation Group, Johns Hopkins University, Lawrence Berkeley National Laboratory, Max Planck Institute for Astrophysics, Max Planck Institute for Extraterrestrial Physics, New Mexico State University, New York University, Ohio State University, Pennsylvania State University, University of Portsmouth, Princeton University, the Spanish Participation Group, University of Tokyo, University of Utah, Vanderbilt University, University of Virginia, University of Washington, and Yale University.

This research has made use of the NASA Astrophysics Data System (ADS) Bibliographic Services and the NASA/IPAC Infrared Science Archive (IRSA), which is funded by the National Aeronautics and Space Ad- ministration and operated by the California Institute of Technology.

We appreciate the expert assistance of the staff at the various observatories where data were obtained.     

\end{acknowledgments}

\facilities{ADS, Bok (B\&C), Gemini (GMOS, GNIRS), IRTF (SpeX), LBT (MODS), LCOGT (FLOYDS, Sinistro), MMT (Binospec, Blue Channel, MMIRS), Keck I (LRIS, NIRES), Shane (Kast), JWST (NIRSpec, MIRI).}

\software{Astropy (\citealt{astropy_2018}), BANZAI (\citealt{McCully_2018}), Binospec pipeline (\citealt{Kansky_2019}), DRAGONS (\citealt{Labrie_2023}), FLOYDS pipeline (\citealt{Valenti_2014}), IRAF (\citealt{Tody_1986}), lcogtsnpipe (\citealt{Valenti_2016}), Light Curve Fitting (\citealt{Hosseinzadeh_2023_lc}), LPipe (\citealt{Perley_2019}), Matplotlib (\citealt{Hunter_2007}), modsCCDred (\citealt{Pogge_2019a,Pogge_2019b}), NumPy (\citealt{Oliphant_2006}), PyRAF (\citetalias{Pyraf_2012}), SciPy (\citealt{Virtanen_2020}), specutils (\citealt{specutils_2025}), Spextool (\citealt{Cushing_2004}), WISeREP (\citealt{wiserep_2012}}

\startlongtable
\begin{deluxetable*}{lcccccc}
\tablecaption{Log of optical spectroscopy of SN~2023ixf \label{tab:opt_spec_log}}
\tablehead{
\colhead{Date} & 
\colhead{MJD} &
\colhead{Phase} &
\colhead{Telescope} &
\colhead{Instrument} &
\colhead{Exp.~Time}\\[-7pt]
\colhead{(UTC)} & 
\colhead{(J2000)} & 
\colhead{(days)} &
\colhead{} &
\colhead{} &
\colhead{(s)}}
\startdata
2023-08-15 & 60171.30 & \phn88.51 & FTN & FLOYDS & 300 \\  
2023-08-20 & 60176.26 & \phn93.47 & FTN & FLOYDS & 600 \\   
2023-08-22 & 60178.25 & \phn95.46 & FTN & FLOYDS & 600 \\  
2023-08-26 & 60182.24 & \phn99.45 & FTN & FLOYDS & 600 \\  
2023-08-30 & 60186.23 & 103.44 & FTN & FLOYDS & 600 \\  
2023-09-05 & 60192.23 & 109.44 & FTN & FLOYDS & 600 \\  
2023-09-07 & 60194.11 & 111.32 & MMT & Binospec & $200\times6$ \\
2023-09-09 & 60196.23 & 113.44 & FTN & FLOYDS & 600 \\  
2023-10-04 & 60221.10 & 138.31 & MMT & Blue Channel & $180\times5$ \\
2023-10-05 & 60222.12 & 139.33 & MMT & Blue Channel & $120\times3$ \\
2023-12-03 & 60281.61 & 198.82 & FTN & FLOYDS & 2700 \\  
2023-12-07 & 60285.51 & 202.72 & MMT & Blue Channel & $600\times4$ \\
2023-12-12 & 60290.64 & 207.85 & Keck & LRIS & $150\times2$ \\
2023-12-13 & 60291.49 & 208.70 & MMT & Binospec & $600\times4$ \\
2023-12-14 & 60292.47 & 209.68 & Bok & B\&C & $1500\times4$ \\ 
2023-12-15 & 60293.51 & 210.72 & LBT & MODS & $400\times7\times2$ \\ 
2023-12-27 & 60305.60 & 222.81 & FTN & FLOYDS & 2700 \\  
2024-01-10 & 60319.51 & 236.72 & MMT & Binospec & $400\times5$ \\
2024-01-19 & 60328.52 & 245.73 & MMT & Binospec & $600\times4$ \\
2024-01-21 & 60330.49 & 247.70 & FTN & FLOYDS & 2700 \\  
2024-02-07 & 60347.44 & 264.65 & FTN & FLOYDS & 2700 \\  
2024-02-14 & 60354.31 & 271.52 & Bok & B\&C & $1500\times4$ \\ 
2024-02-28 & 60368.49 & 285.70 & FTN & FLOYDS & 2700 \\  
2024-03-03 & 60372.31 & 289.52 & MMT & Binospec & $600\times4$ \\
2024-03-03 & 60372.44 & 289.65 & Bok & B\&C & $1500\times5$ \\ 
2024-03-10 & 60379.50 & 296.71 & LBT & MODS & $600\times4\times2$ \\ 
2024-03-19 & 60388.38 & 305.59 & FTN & FLOYDS & 2700 \\  
2024-03-20 & 60389.50 & 306.71 & Shane & Kast & $1200\times3$ \\
2024-04-05 & 60405.48 & 322.69 & MMT & Binospec & $600\times4$ \\
2024-04-06 & 60406.58 & 323.79 & Keck & LRIS & $600\times4$ \\
2024-04-10 & 60410.57 & 327.78 & FTN & FLOYDS & 1800 \\  
2024-04-14 & 60414.37 & 331.58 & Bok & B\&C & $1500\times5$ \\ 
2024-04-19 & 60419.45 & 336.66 & Shane & Kast & $1200\times4$ \\
2024-05-01 & 60431.53 & 348.74 & Keck & LRIS & 300 \\
2024-05-10 & 60440.40 & 357.61 & Bok & B\&C & $1500\times4$ \\ 
2024-05-11 & 60441.41& 358.62 & LBT & MODS & $600\times4\times2$ \\  
2024-05-25 & 60455.40 & 372.61 & FTN & FLOYDS & 1800 \\  
2024-05-30 & 60460.35 & 377.56 & LBT & MODS & $600\times4\times2$ \\
2024-06-11 & 60472.45 & 389.66 & FTN & FLOYDS & 1800 \\  
2024-06-29 & 60490.38 & 407.59 & FTN & FLOYDS & 3600 \\  
2024-08-01 & 60523.30 & 440.51 & FTN & FLOYDS & 3600 \\  
2024-08-08 & 60530.22 & 447.43 & Shane & Kast & $1200\times5$\\
2024-12-11 & 60655.48 & 572.69 & LBT & MODS & $900\times4\times2$ \\ 
2024-12-16 & 60660.46 & 577.67 & LBT & MODS & $900\times6\times2$ \\ 
2025-02-02 & 60708.46 & 625.67 & LBT & MODS & $900\times7\times2$ \\ 
2025-02-05 & 60711.46 & 628.67 & MMT & Binospec & $900\times6$ \\
2025-02-20 & 60726.43 & 643.64 & MMT & Blue Channel & $1800\times3$ \\
2025-02-22 & 60728.44 & 645.65 & MMT & Blue Channel & $1800\times4$ \\
2025-04-27 & 60792.32 & 709.53 & MMT & Binospec & $900\times6$ \\
2025-05-17 & 60813.00 & 730.21 & Gemini-N & GMOS & $1200\times3$ \\
2025-05-26 & 60821.34 & 738.55 & LBT & MODS & $1200\times5\times2$ \\ 
2025-06-05 & 60831.35 & 748.76 & MMT & Blue Channel & $1800\times3$ \\[+2pt]
\enddata
\end{deluxetable*}

\begin{deluxetable*}{lcccccc}
\tablecaption{Log of infrared spectroscopy of SN~2023ixf\label{tab:nir_spec_log}}
\tablehead{
\colhead{Date} & 
\colhead{MJD} &
\colhead{Phase} &
\colhead{Telescope} &
\colhead{Instrument} &
\colhead{Exp.~Time}\\[-7pt]
\colhead{(UTC)} & 
\colhead{(J2000)} & 
\colhead{(days)} &
\colhead{} &
\colhead{} &
\colhead{(s)}}
\startdata
2023-12-04 & 60282.49 & 199.70 & MMT & MMIRS & $120\times 6$\\
2024-01-31 & 60340.38 & 257.59 & MMT & MMIRS & $120\times 7$\\
2024-02-21 & 60361.52 & 278.73 & IRTF & SpeX & $200\times 8$\\
2024-03-23 & 60392.55 & 309.76 & IRTF & SpeX & $300\times 12$\\
2024-05-25 & 60455.37 & 372.58 & MMT & MMIRS & $120\times 12$\\
2024-07-09 & 60500.35 & 417.56 & IRTF & SpeX\tablenotemark{$\rm *$} & $120\times 16$\\
2024-08-12 & 60534.28 & 451.49 & IRTF & SpeX\tablenotemark{$\rm *$} & $120\times 24$\\[+2pt]
\enddata
\tablecomments{The $+200$ d and $+258$ d MMT/MMIRS spectra were previously published by \cite{Park_2025}.}
\tablenotetext{$\rm *$}{Denotes observations taken in PRISM mode of IRTF/SpeX.}
\end{deluxetable*}

\appendix

\restartappendixnumbering

\section{Effects of MIRI/LRS Resolution on Double-peaked Profiles}
\label{app:miri_res}

\begin{figure*}[t!]
    \centering
    \includegraphics[width=\textwidth]{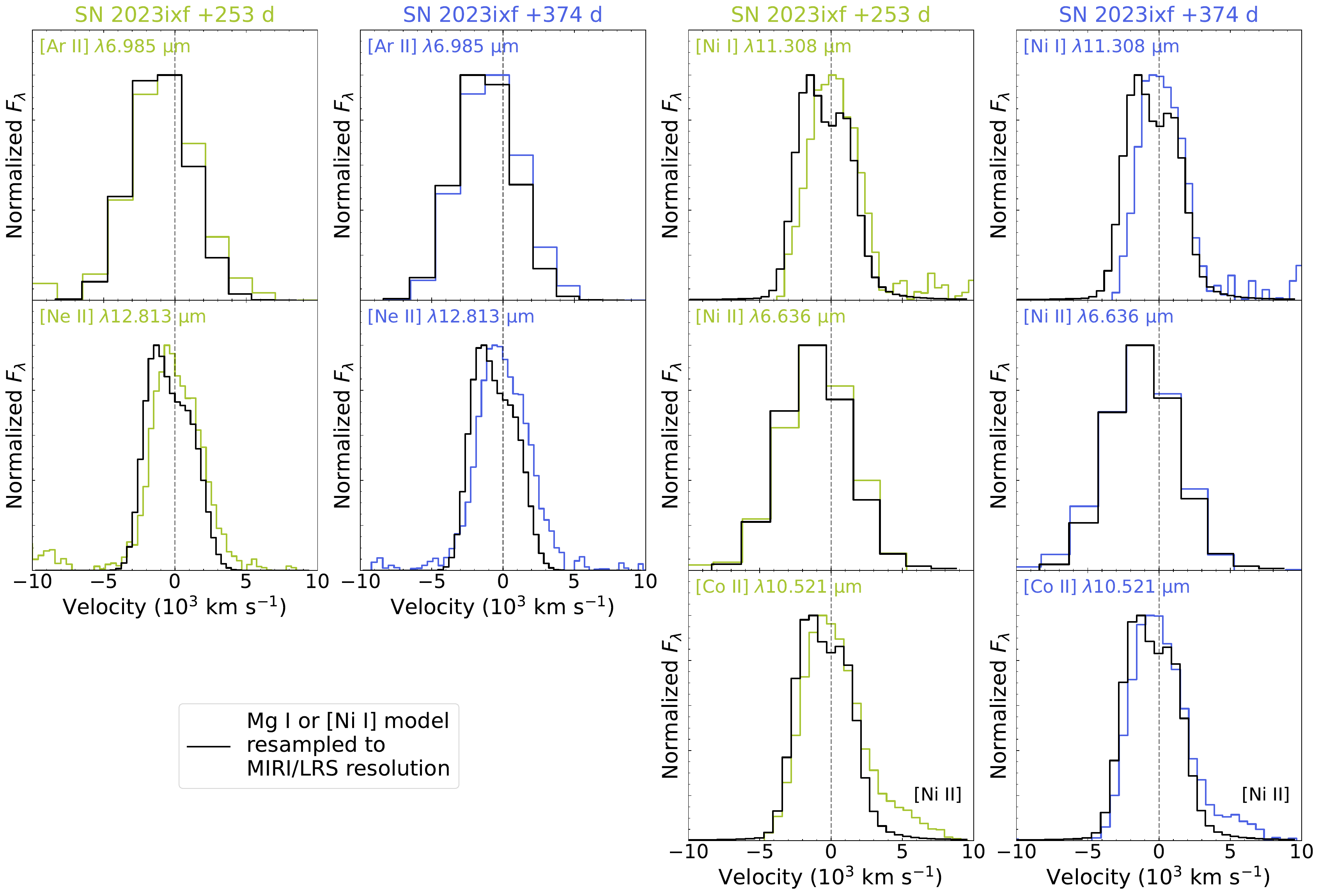}
    \caption{The effects of MIRI/LRS on double-peaked profiles. We smooth and resample template double-peaked profiles (best-fit \ion{Mg}{1} $1.504\ {\rm \mu m}$ model for IMEs and best-fit [\ion{Ni}{1}] $3.12\ {\rm \mu m}$ model for IGEs) to match the resolution of MIRI/LRS for each line considered. The low spectral resolution of MIRI/LRS can only marginally resolve the characteristic peak separation of $\approx 2800\ {\rm km\ s^{-1}}$ in intrinsically double-peaked emission lines at the longest wavelengths ($>10\ {\rm \mu m}$).
    With existing data, we are unable to disentangle resolution effects from any intrinsically double-peaked profiles in the MIRI/LRS observations.
    }
    \label{fig:miri_resampled}
\end{figure*}

In this Appendix, we investigate whether the broad, symmetric line profiles from predominantly singly ionized IMEs and IGEs --- with the sole exception of [\ion{Ni}{1}] $11.308\ {\rm \mu m}$, which arises from neutral stable nickel --- are due to resolution effects or intrinsic geometric distributions.
Specifically, we test the effects of resolution on our ability to identify any potential double peaks in these longer-wavelength MIR lines.
For each line, we smooth and resample a double-peaked profile template to match the observed resolution at the central wavelength of the line under consideration. 
We use the best-fit [\ion{Ni}{1}] $3.12\ {\rm \mu m}$ model for [\ion{Ni}{1}] $11.308\ {\rm \mu m}$, [\ion{Ni}{2}] $6.636\ {\rm \mu m}$, and [\ion{Co}{2}] $10.521\ {\rm \mu m}$, and the best-fit \ion{Mg}{1} $1.504\ {\rm \mu m}$ model for [\ion{Ar}{1}] $6.985\ {\rm \mu m}$ and [\ion{Ne}{2}] $12.813\ {\rm \mu m}$.
As shown in Figure~\ref{fig:miri_resampled}, an intrinsically double-peaked profile with a characteristic peak separation of $\sim 2800\ {\rm km\ s^{-1}}$ is only marginally resolvable at the lower resolution of MIRI/LRS.
Given the likely differing physical conditions (e.g., a lower density or a higher temperature; \citealt{Dessart_2025b,Jacobson-Galan_2026}) required for these lines from mostly singly ionized IMEs and IGEs, it is possible that they have a smaller peak separation, which becomes indistinguishable at a lower resolution.

\clearpage

\bibliography{Reference}{}
\bibliographystyle{aasjournalv7}
\end{document}